\documentclass[12pt]{emulateapj}
\usepackage{epsfig}
\usepackage{amsmath}
\usepackage{amssymb}
\usepackage{natbib}
\usepackage{graphicx}
\usepackage{enumerate}
\usepackage{algorithm}
\usepackage{algpseudocode}

\shorttitle{Post-Newtonian SMBH dynamics in GADGET-3}
\shortauthors{Rantala et al.}

\newcommand{\etabs}{\ensuremath{\eta_{\mathrm{GBS}}}}

\newcommand{\archain}{AR-CHAIN}
\newcommand{\ketju}{KETJU}
\newcommand{\gadget}{GADGET-3}
\newcommand{\norm}[1]{\|#1\|}
\newcommand{\abs}[1]{\left|#1\right|}
\newcommand{\ud}{\mathrm{d}}

\DeclareMathOperator{\sgn}{sign}
\newcommand{\vect}[1]{\boldsymbol{#1}}

\newcommand{\derfrac}[2]{\frac{\ud #1}{\ud #2}}

\newcommand{\chdrift}{\mathbf{X}}
\newcommand{\chkick}{\mathbf{V}}
\newcommand{\Rsun}{R_\odot}
\newcommand{\Msun}{M_\odot}

\newcommand{\fromto}{\rightarrow} 
\newcommand{\bigO}{\mathcal{O}} 

\newcommand{\dttree}{\Delta t_\text{tree}}

\begin{document}

\title{Post-Newtonian dynamical modeling of supermassive black holes in 
galactic-scale simulations}

\author{Antti Rantala$^1$, Pauli Pihajoki$^1$, Peter H. Johansson$^1$, 
Thorsten Naab$^2$, Natalia Lah\'en$^1$, Till Sawala$^1$}

\affil{$^1$ Department of Physics, University of Helsinki, Gustaf 
H\"allstr\"omin katu 2a, Finland;\\}
\affil{$^2$ Max-Planck-Insitut f\"ur Astrophysik, Karl-Schwarzchild-Str. 1, 
D-85748, Garching, Germany; }

\email{antti.rantala@helsinki.fi}

\begin{abstract}

We present \ketju{}, a new extension of the widely-used 
smoothed particle hydrodynamics simulation code \gadget{}. The key feature of 
the code is
the inclusion of algorithmically regularized regions around every 
supermassive black hole 
(SMBH). This allows for simultaneously following global galactic-scale 
dynamical and astrophysical processes, while solving the dynamics of SMBHs, 
SMBH binaries 
and surrounding stellar systems at sub-parsec scales. The \ketju{} code 
includes Post-Newtonian terms in the 
equations of motions of the SMBHs which 
enables a new SMBH merger criterion based on the gravitational wave coalescence 
timescale pushing the merger separation of SMBHs down to $\sim 0.005 \ \rm pc$.
We test the performance of our code by comparison to NBODY7 and rVINE. We set 
up 
dynamically stable multi-component merger progenitor galaxies to study the SMBH 
binary evolution 
during galaxy mergers. In our simulation sample the SMBH binaries do not suffer 
from the final-parsec problem, 
which we attribute to the non-spherical shape of the merger 
remnants. For bulge-only models, the hardening rate decreases with 
increasing resolution, whereas for models which in addition include massive 
dark 
matter halos 
the SMBH binary hardening rate becomes practically independent of the mass 
resolution of the 
stellar bulge. The SMBHs coalesce on average 200 Myr after the formation of the 
SMBH binary. 
However, small differences in 
the initial SMBH binary eccentricities can result in large differences in the 
SMBH 
coalescence times. Finally, we discuss the future prospects of KETJU, which 
allows for a 
straightforward inclusion of gas physics in the simulations.
\end{abstract}

\section{Introduction}

There is ubiquitous evidence for the presence of supermassive black holes 
(SMBHs) with masses in the range of $M=10^{6}-10^{10} M_{\odot}$ in the centers 
of all 
massive galaxies in the local Universe (e.g. 
\citealt{Kormendy1995,Ferrarese2005,kormendy2013}). In addition, there is 
a strong suggestion of a co-evolution of SMBHs and their host galaxies 
as manifested in the surprisingly tight relations between the SMBH masses and 
the 
fundamental 
properties of the galactic bulges that host them, e.g. the bulge mass 
\citep{Magorrian1998,Haring2004} and the bulge stellar velocity dispersion 
\citep{Gebhardt2000,Ferrarese2000,Tremaine2002}.  

In the hierarchical picture of structure formation, galaxies grow bottom-up 
through mergers and gas accretion \citep{1978White}. Massive, slowly-rotating 
early-type galaxies, that are expected to host the largest SMBHs in the 
Universe, are believed to have assembled through a two-stage process. The early 
assembly is dominated by rapid in situ star formation fueled by cold gas flows 
and hierarchical merging of multiple star-bursting progenitors, whereas the 
later 
growth below redshifts 
of $z\lesssim 2-3$ is dominated by a more quiescent phase
of accretion of stars formed mainly in progenitors outside the main galaxy 
(e.g. 
\citealt{2009Naab,oser2010,2011Feldmann,2012Johansson,2015Wellons,
rodriguez-gomez2016,qu2016}). See also \cite{Naab2016} for a review.

This hierarchical formation process will result in situations with 
multiple SMBHs in the same stellar system (e.g. 
\citealt{begelman1980, volonteri2003}).
These SMBHs will subsequently sink to the center of the galaxy due to dynamical 
friction from stars and gas and form a wide gravitationally bound binary with a 
semi-major axis of $a \sim 10$ 
pc. Next, the semi-major-axis of the binary will 
shrink ('harden') due to 
the interaction of the binary with the stellar background. A star crossing 
at a distance of the order of the semi-major axis of the binary will experience 
a complex three-body 
interaction with the binary and carry away energy and angular momentum from the 
SMBH binary 
system 
(eg. \citealt{hills1980}). If the population of stars with centrophilic orbits 
is not depleted, the binary will harden at an approximately constant rate:
\begin{equation}
\derfrac{}{t} \left( 
\frac{1}{a} \right) \propto \frac{G \rho_\star}{\sigma_\star},
\end{equation}
assuming a constant stellar density $(\rho_\star)$ and velocity dispersion 
$(\sigma_\star)$ at the center of the galaxy \citep{quinlan1996}. 
If the center-crossing (or 'SMBH loss cone') orbital population is depleted, 
the binary hardening is halted. This so-called final-parsec problem is 
persistently present in simulations of isolated collisionless spherically 
symmetric stellar systems \citep{Milosavljevic2001,Milosavljevic2003}.

Recent numerical work suggests that the problem 
is less severe or might even be nonexistent in simulations of triaxial 
\citep{berczik2006} and axisymmetric galaxies \citep{khan2013}, for which the 
added asymmetric perturbations in the gravitational potential are able to 
refill 
the loss cone 
by repopulating centrophilic stellar orbits. Similarly, the merger of two 
galaxies 
will break the symmetry of the galactic potentials resulting in a more 
efficient 
refilling of the loss 
cone and thus potentially avoiding the final-parsec problem 
\citep{Preto2011,khan2011,Khan2012}. However, even in simulations that avoid 
the 
final-parsec problem the 
loss-cone filling rate is affected by the enhanced two-body 
relaxation timescale, especially in simulations with $N \lesssim 10^6$ 
particles \citep{Vasiliev2015}.
Recently, \citet{gualandris2016} also studied the collisionless loss-cone 
repopulation in 
triaxial galaxies without 
SMBHs using an accurate fast multipole method and found that for particle 
numbers $N<10^7$, the loss-cone filling rate is mildly $N$-dependent, whereas 
the rate 
becomes practically independent of $N$ for particle numbers above $N\sim 10^7$. 

Recent observations show that even early-type galaxies have non-negligible gas 
reservoirs of cold gas in their central regions \citep{young2011, davis2013}. 
The inclusion of gas in the simulations tends to result in 
non-axisymmetric gas torques and a remnant that is denser in the central 
regions due to the dissipative nature of the gaseous component. Both of these 
effects 
facilitate rapid hardening of a SMBH binary 
and might help in solving the final-parsec problem 
\citep{Armitage2005,Mayer2007}. 
This is especially true at high redshifts where very gas-rich mergers are 
expected to occur frequently \citep{Khan2016}. Indeed, there is observational 
evidence for the presence of massive black holes from strong nuclear outflows 
at 
$z \sim 1-2$ \citep{genzel2014}. However, the results from 
hydrodynamical simulations depend sensitively on the adopted feedback and star 
formation model, and thus we caution that it is not yet clear, whether the 
inclusion of gaseous component on its own is sufficient for solving the 
final-parsec problem (see e.g. \citealt{Chapon2013}).

If the final-parsec problem is avoided, the loss of orbital energy eventually 
becomes dominated by the emission of gravitational waves at very small 
centiparsec 
binary separations with a strong dependence on the binary eccentricity. 
Recently, this process was spectacularly confirmed by the direct detection of 
gravitational 
waves from a stellar mass BH binary system by \citet{Abbott2016}. Future 
space-borne 
low-frequency laser interferometers are expected to detect
a similar signal from supermassive black hole binary systems (e.g. 
\citealt{Amaro-Seoane2012}).
 
To model the dynamics of SMBHs in galaxy mergers, one would ideally run a 
simulation that simultaneously resolves the global kpc-scale dynamics of 
the merger and the subparsec evolution of the SMBH binary. This is a very 
ambitious goal and typically only one of these scales has been properly 
resolved and modeled in any given simulation. In the past decade there has 
been 
significant progress in simulating both galaxy mergers and the full 
cosmological evolution of galaxies including the effects of SMBHs initially 
using smoothed 
particle hydrodynamics (SPH) 
(e.g.\citealt{DiMatteo2005,springel2005,Sijacki2007,Johansson2009b,
Johansson2009a,Booth2009,Choi2012}) and later also using both 
adaptive-mesh refinement (AMR) (e.g. \citealt{Kim2011,Martizzi2012,Dubois2012}) 
and moving mesh techniques (e.g. 
\citealt{Vogelsberger2013,Costa2014,Sijacki2015,kelley2017,kelley2017b}). These
simulations allow for a large number of particles and are very successful in 
capturing the global structure of gas and stars in the galaxies. In addition, 
they are able to approximatively follow additional astrophysical
processes by including sophisticated subresolution models for gas cooling, 
star formation, metal enrichment and the feedback from SMBHs and evolving 
stellar populations.

However, the fundamental limitation of this approach is that only a limited 
number of particles or grid cells sample the underlying smooth gravitational 
potential and by necessity the gravitational force must be softened to reduce 
the graininess of the potential. 
The softening length or equivalently the size of the minimum grid cell
sets a natural resolution limit, below which the dynamics, such as the close 
two-body encounters with a massive SMBH cannot be modeled accurately. This also
leads to uncertainties in dynamical friction timescales of SMBHs. 
A possibility that circumvents this problem is the addition of a 
subresolution 
drag force term into the equations of motion of the sinking SMBH that accounts
for the  unresolved encounters of the SMBH and the field stars. This method 
is particularly well suited for 
cosmological simulations, which typically have low spatial resolution 
\citep{Tremmel2015}.

Gravitational softening will also affect the density and velocity profiles of 
stars in the 
centers of galaxies, which strongly interact with the SMBHs. Finally the 
hardening and 
merging timescales of binary SMBHs are also plagued by large uncertainties
and the common 'a priori' assumption often taken in these models is that both 
the hardening and merging of SMBHs happens rapidly, with the actual 
implementation then proceeding through a subresolution model with limited 
predictive power. 
 
An alternative approach is to calculate the gravitational force directly by 
summing exactly the force from every particle on every particle. This method
is computationally expensive, but allows in combination with high-order 
integrators for a very accurate calculation of the gravitational forces. It is 
widely used to simulate collisional N-body systems (e.g. 
\citealt{Aarseth1999}). 
This method does not require gravitational softening but the computational time 
scales steeply with the particle number $\bigO(N^2)$ as opposed to tree and 
mesh codes, which typically scale as $\bigO(N \log N)$. In addition it is not 
straightforward to 
model the gaseous component present in galaxies in a pure N-body code and 
combined with 
the limited number of particles ($N\sim 10^{6}$, \citealt{Wang2016}) this 
limits 
the 
applicability of these codes for a self-consistent treatment of SMBH dynamics 
in 
a full 
galactic environment. Thus, owing to these inherent limitations current 
numerical simulations with N-body codes have typically only explored separate 
aspects of the full 
problem by limiting themselves to studies of SMBH binary dynamics in the centers
of isolated galaxies or merger remnants, with the surrounding galaxy often 
represented by idealized initial conditions (e.g. 
\citealt{Milosavljevic2001,Milosavljevic2003,berczik2006,Preto2011,khan2011,
khan2013,Gualandris2012,Vasiliev2014b,Holley_Bockelmann2015}). 
An important distinction to also keep in mind is the difference between 
the force accuracy of a simulation and the actual simulation accuracy that also 
depends on how accurately the orbits of the particles can be integrated. A 
recent paper 
by \citet{Dehnen2004}
demonstrated that a suitably tuned fast multipole method is capable of 
delivering a force accuracy 
comparable to that of a direct-summation code, while still retaining a very 
efficient
$\bigO(N)$ scaling. However, for our purposes in addition to an accurate force 
calculation, it is also 
of paramount importance that we are able to accurately integrate the equations 
of motion without softening 
and to resolve arbitrarily close encounters in the vicinity of the SMBHs.

The main goal of this article is to present and test our new code that 
attempts to combine the best aspects of the two numerical approaches. Our code 
\ketju \
(the word for 'chain' in Finnish) combines an algorithmic chain regularization 
(\archain)
method to efficiently and accurately compute the dynamics close to SMBHs with 
the fast and widely used tree code \gadget \  \citep{springel2005b} for the 
calculation of the global galactic dynamics. 
The performance of normal \gadget \ can be substandard in situations 
that require high dynamical precision due to the insufficient precision of the 
tree 
force calculation 
(see e.g. \citealt{gualandris2016}). Some of these problems can be mitigated by 
setting the internal
accuracy parameters of \gadget \ to very high values, significantly beyond 
their 
usual nominal values. In 
addition in our \ketju \ code the strongest interactions between particles will 
be resolved within
the regularized algorithmic chain region, and not treated by standard \gadget. 
The main advantage of building
\ketju \ on the \gadget \ platform is that it enables the use of a rich set of 
astrophysical cooling and feedback 
models for future \ketju \ runs that also include a gaseous component.

A similar hybrid approach of combining a tree code with a regularization 
algorithm was originally implemented by \cite{Jernigan1989} and 
\cite{McMillan1993}.   
\cite{Oshino2011} and 
\cite{Iwasawa2015} also both combined the tree algorithm with a direct 
summation 
code, however without the inclusion of regularization, whereas the BRIDGE 
framework developed 
by \cite{Fujii2007} allows for the combination of different types of N-body 
codes. For our purposes the most 
relevant precursor code is the rVINE code \citep{karl2015}, which is very 
similar in spirit and functionality to our 
code. In rVINE an algorithmically regularized region is inserted around a 
single SMBH with this structure embedded in the VINE code, which is an 
OpenMP-parallelized tree/SPH 
code employing a binary tree algorithm \citep{Wetzstein2009,Nelson2009}. 
Although similar to rVINE, there are significant
differences and improvements in the \ketju\ code detailed in \S 
\ref{regularized_Gadget} and \S \ref{ARCHAIN_int}. As opposed to rVINE, 
the \ketju\ code allows for multiple regularized chains with an individual 
chain system for each SMBH particle. In addition, the \ketju\ code includes 
Post-Newtonian (PN) correction 
terms up to order PN3.5, which in principle allows for accurate dynamics valid 
down to $\sim 10$ Schwarzschild radii.  

We begin this article by covering the structure of the chain subsystems in \S 
\ref{regularized_Gadget} and describe how they are integrated into the 
\gadget{} code. Next, we present the properties of the algorithmic chain 
regularization 
method in \S \ref{ARCHAIN_int}. The intricate details of the algorithms used in 
the 
code are discussed in Appendices \ref{sc:algoreg} and \ref{appendix_B}.
In \S \ref{Test_code} we test and calibrate our code against rVINE and the 
direct summation code NBODY7. The 
performance and scalability of the code are discussed in \S \ref{performance}. 
In \S 
\ref{results} we use the \ketju \ code to study the 
resolution dependence of the SMBH hardening rates in merger simulations of both 
two- and three-component galaxy models. We discuss our results and plans for 
the 
future 
in \S \ref{discussion} and finally present our conclusions in \S 
\ref{conclusions}.

\section{Regularized subsystems in Gadget-3}
\label{regularized_Gadget}

\subsection{The chain subsystem}
\label{chain_subsystem}
In the standard \gadget{} code \citep{springel2005b} the gravitational 
accelerations 
of N-body particles are computed 
using either a pure tree algorithm or a hybrid tree-mesh TreePM algorithm. In 
the TreePM method, the gravitational tree is used to compute the short-range 
forces while an efficient particle mesh method is used for the long-range 
component.  Hereafter, these two force calculation procedures are referred 
to simply as the tree method. 

\gadget{} propagates simulation particles using a symplectic 
kick-drift-kick (KDK) leapfrog scheme with individual adaptive timesteps 
\citep{springel2005b}. 
To integrate the regularized \archain{} algorithm in the \gadget{} code, we 
must 
first select a subset of particles from the complete set of simulation 
particles. This
regularized subset of particles is excluded from the \gadget{} tree force 
calculation and the standard leapfrog propagation. These particles are instead 
propagated using the \archain{} integrator of \ketju{} (hereafter, chain 
integration). The chain integration procedure is presented thoroughly in the 
next section.

We divide the simulation particles into three categories according to their type
and distance to SMBH particles:
\begin{enumerate}
 \item Chain particles: all the SMBH particles and the stellar particles which
lie in the immediate vicinity of a SMBH particle. The SMBH and the
surrounding stellar particles form a chain subsystem. Note that in the current 
implementation 
gas and dark matter particles cannot enter the chain subsystems.
 \item Perturber particles: all the simulation particles which induce a strong 
tidal perturbation on a chain subsystem. These particles feel the tree
force and are propagated using the \gadget{} leapfrog, but in addition they 
receive a correction to their accelerations from a nearby chain subsystem.
 \item Tree particles: simulation particles that are far away from any of the 
SMBH particles and are thus treated as ordinary \gadget{} particles with 
respect to the force calculation.
\end{enumerate}

We have implemented a distance-based selection criterion for chain 
subsystem members, in which the chain radius of a SMBH 
($\bullet$) depends on the mass of the SMBH: 
\begin{equation}\label{rinfl}
\frac{r_{\mathrm{chain}}}{1\text{ kpc}} = \lambda \times 
\frac{M_{\bullet}}{10^{10} M_\odot},
\end{equation}
where $\lambda$ is a user specified dimensionless input parameter. Note that 
using this definition the 
chain radius of a SMBH remains constant in simulations with no black hole mass 
accretion or mergers.

\begin{figure}[h!]
\centering
\includegraphics[width=\linewidth]{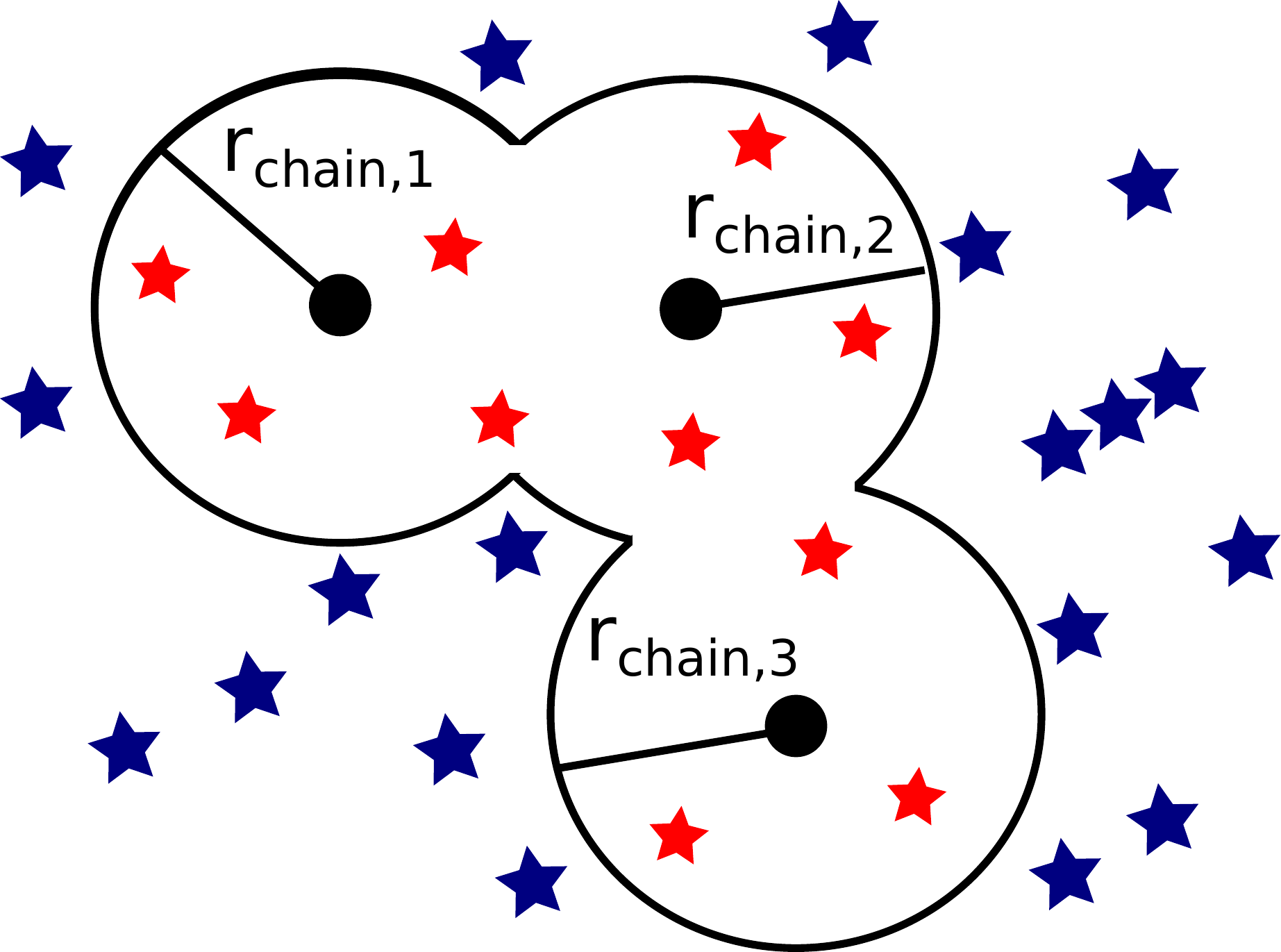}
\caption{An illustration of the chain subsystem with three SMBHs (black filled 
circles) and
ten stellar chain members (red stars). The blue stars outside the SMBHs'
chain radii act as perturbing particles.}
\label{makkara}
\end{figure}

After the radius of a SMBH has been set, we select the members 
of a chain subsystem. A stellar particle ($\star$) belongs to a chain subsystem 
of a SMBH if
\begin{equation}
 \norm{\vect{r}_{\star} - \vect{r}_{\bullet}} < 
r_{\mathrm{chain}}.
\end{equation}
For a schematic illustration of a chain subsystem among tree
particles, see Fig. \ref{makkara}.

The individual chain particles are removed from the tree force calculation. 
However, the center-of-mass (CoM) of the chain subsystem is placed into the 
tree 
structure as a 'macro' particle, with the combined mass of all its chain 
particles. This macro particle acts as an ordinary collisionless tree particle 
in the simulation. 

The macro particle is an ordinary \gadget{} tree particle and must have 
a non-zero gravitational softening length. \gadget{} uses as a  
gravitational softening kernel the  Monaghan-Lattanzio spline kernel 
\citep{monaghan1985}, which is exactly 
Newtonian outside the softening length $h_{\mathrm{ML}}$. The usually quoted 
Plummer-equivalent softening length $\epsilon$, which is used to set the 
softening
lengths of simulation particles, is related to this quantity through 
$h_{\mathrm{ML}} = 2.8 \times \epsilon$ \citep{Springel2001}. Thus, we enforce 
the condition 
\begin{equation}
 r_{\mathrm{chain}} > 2.8 \times \epsilon
\end{equation}
for the chain radius and the Plummer-equivalent gravitational softening length 
in order to ensure that the mutual gravitational interactions of stars and 
SMBHs are never softened in \ketju.

The chain subsystems are initialized at the beginning of the 
simulation, after which the status of chain particles is updated after every 
chain integration interval. If, at the next timestep, a chain particle has 
propagated 
outside of the chain radius of a SMBH, an escape event occurs and the chain 
particle is 
restored to the tree. If there are multiple SMBHs in a single chain, we check 
the escape
condition for all chain particle - SMBH pairs separately. The absorption of new 
particles into a chain subsystem is performed similarly. An absorption 
event occurs if a tree particle enters the chain radius of a SMBH. The 
center-of-mass properties and the total mass of the macro particles are updated 
after both absorption and escape events. We terminate a chain subsystem if the 
SMBH is the only remaining chain member particle, i.e. $N_{\mathrm{c}} < 
2$. A new active chain subsystem is initialized if stellar particles are found 
inside the chain radius 
of a SMBH.

\subsection{Tidal perturbations and force corrections}\label{perturbers}
In addition to internal forces, the dynamics of the particles in a chain 
subsystem is affected by external
gravitational forces $\vect{f}_\mathrm{i}$. These external forces are dominated 
by the
closest neighboring tree particles which are not chain subsystem members. We 
define and select these perturber 
particles of a subsystem with a tidal criterion before every chain integration. 
A tree 
particle, labeled with index
$j$, is selected as a perturber of a chain subsystem if the condition

\begin{equation}
\begin{aligned}
\norm{\vect{r}_{j} - \vect{r}_{\bullet}} < \gamma 
\times
r_{\mathrm{chain}}\left(\frac{m_{\mathrm{j}}}{M_{\bullet}}\right)^{1/3} = 
r_{\mathrm{pert, j}}
\label{tidalgamma}
\end{aligned}
\end{equation}
is satisfied with any of the SMBHs in the particular chain subsystem. Here
$\gamma$ is a user-defined tidal parameter and $r_\mathrm{chain}$ is the SMBH 
chain
radius defined in Eq. \eqref{rinfl}. In the case of equal-mass perturber 
particles, the perturber radius is identical for all the particle species. 
The tidal parameter $\gamma$ is chosen as such that the perturber radius 
is a few times the chain radius of the SMBH. For an illustration of a 
perturbed subsystem see Fig. \ref{fig: 
chainpert}. In the case of unequal-mass particles, more massive particles can 
become perturbers even if they lie further away from the SMBH than lighter 
simulation particles and thus there are several perturber radii. The 
user-defined 
parameters in \ketju{} are 
listed in Table \ref{table: params} with their typical values from simulations 
appearing in \S \ref{results}.

\begin{table}
\begin{center}
	 \caption{User-defined \ketju{} parameters with typical values in the 
simulations from \S \ref{results}.}
    \begin{tabular}{| c | c | c|}
    \hline
    Parameter& Description & Numerical value\\ \hline 
    $\lambda$ &Sets the chain radius& 1.8\\ 
    $\gamma$ &Sets the perturber radius& 25\\
    \etabs & \archain{} integrator accuracy & $10^{-6}$\\
    \hline
    \end{tabular}
    \label{table: params}
    \end{center}
\end{table}

\begin{figure}[h!]
\centering
\includegraphics[width=\linewidth]{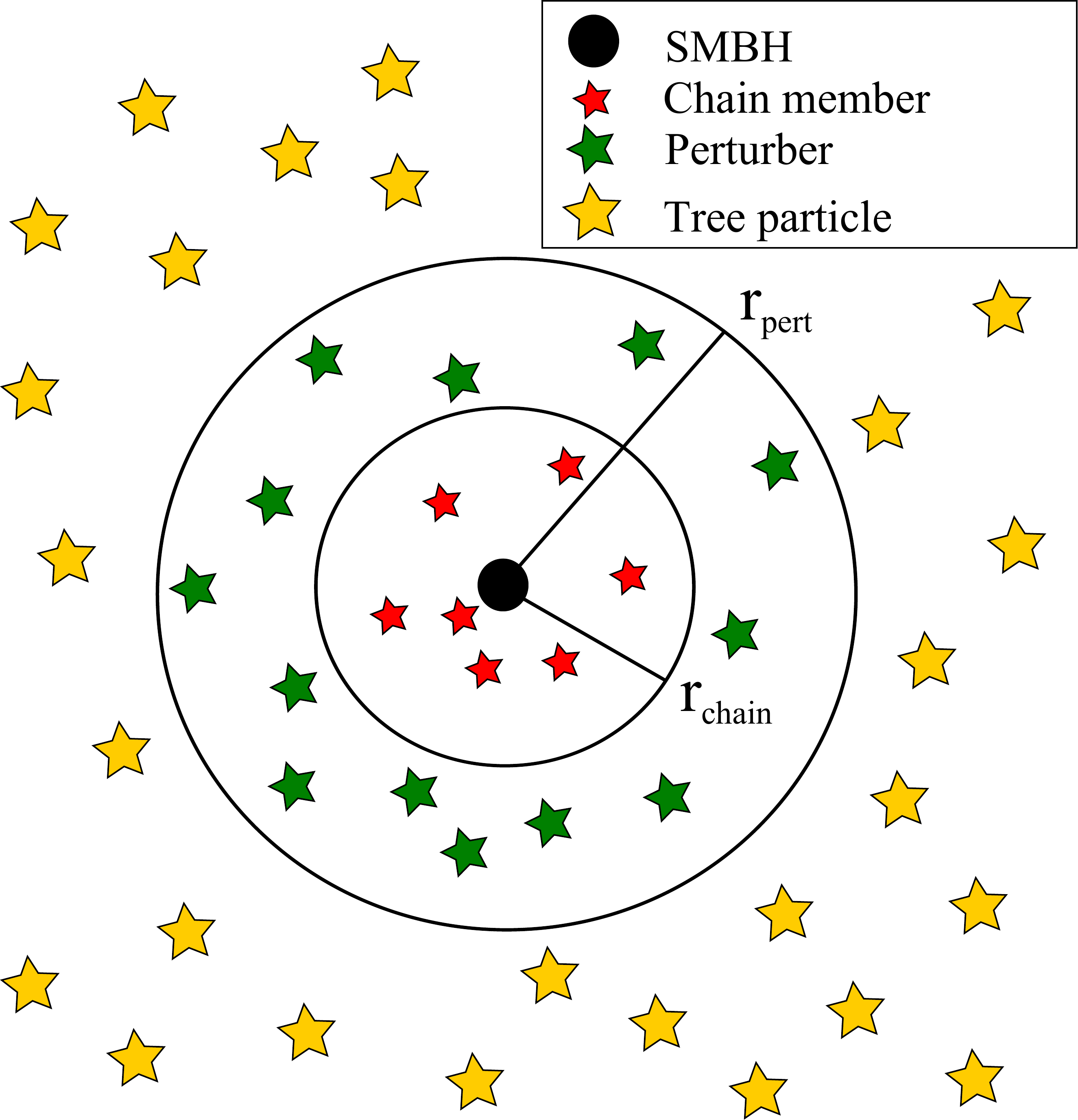}
\caption{An illustration of a chain subsystem with a single SMBH (the black 
filled circle), chain particles (red stars), perturbing tree particles (green 
stars) and
ordinary tree particles (yellow stars). The chain radius of the SMBH is 
marked with $r_{\mathrm{chain}}$ and the radius containing the perturbers with 
$r_{\mathrm{pert}}$. Note
that here we assume equal-mass tree particles so there is a single perturber 
radius.}
\label{fig: chainpert}
\end{figure}

During the chain integration, the external force perturbations 
$\vect{f}_{\mathrm{i}}$ 
are computed using perturber positions, as described in Algorithm
\ref{alg:kick} in Appendix~\ref{sc:chain_leapfrog}. As the \archain{} 
integrator leapfrogs several regularized
substeps during a single tree timestep and the perturber positions are obtained
at the beginning of the timestep, we predict both the positions of the 
macroparticle and the perturber particles using a simple quadratic 
extrapolation. 
In addition, we also include the force contribution of distant tree particles 
as a far-field perturbation which is kept constant during the chain integration.
In general, the tree force calculation does not resolve the subresolution 
dynamics in the chain subsystems. However, in order to satisfy Newton's third 
law, perturber particles receive an extra
force correction from the resolved chain after the completion of the Gadget-3 
tree force calculation. The procedure is as follows. First, we subtract
the contribution of the macro particle from the acceleration of the perturber 
particle. Then, we resolve the positions of the chain particles and compute the 
correction for the acceleration of the perturber using the softened direct 
summation 
method 
employing the Monaghan--Lattanzio cubic spline kernel \citep{monaghan1985} used 
in standard Gadget-3. If a tree particle perturbs multiple chain subsystems, it 
receives a force correction from all of them. In addition, all macro particles 
in the tree receive a force correction due to the perturber forces on the 
resolved individual 
particles in the chain subsystem. To sum up, the internal structure of the 
chain subsystems in visible to the nearby perturbing tree particles. For 
faraway 
tree 
particles the chain subsystems appear as a single macro particles. However, 
this is not a problem since treating distant simulation particles using a 
low-order 
multipole expansion is in fact the essence of the tree force calculation itself 
(e.g. \citealt{barnes1986}).
As the force correction computation scales
proportionally to the number of chain particles and perturber particles, 
$\bigO{(N_{\mathrm{c}} \times N_{\mathrm{p}})}$, the 
perturber tidal parameter $\gamma$ should be selected carefully in order to 
optimize both code accuracy and the resulting computational cost. In 
this paper, we follow the general rule of thumb that
\begin{equation}
r_{\mathrm{pert}} \ge 2 \times r_{\mathrm{chain}},
\end{equation}
i.e. the perturber radius is at least twice the chain radius of the SMBH 
for every particle type in the simulation.

\subsection{Timestepping with chain}\label{timestepping_with_chain}
In \gadget, the timestep of a collisionless particle is set to
\begin{equation}
 \Delta t_{\mathrm{grav}} =  \left( \frac{2 \eta 
\epsilon}{|\vect{a}|}
\right)^{1/2},
\end{equation}

in which $\eta$ is the user-defined error tolerance parameter, $\epsilon$ is 
the 
gravitational softening 
length and
$\vect{a}$ is the acceleration of the particle. In addition, all timesteps 
in \gadget{} are discretized as power of two subdivisions of the global tree 
timestep \citep{springel2005b}.

In \ketju, the timestep criterion is modified slightly. All SMBH particles 
are placed on the smallest active level in the global timestep hierarchy. In 
addition, all the particles escaping from the chain are set to the smallest 
tree timestep level. The chain time integration is performed within \gadget's 
KDK integration cycle in the following way. The subsystem/tree memberships of 
the simulation 
particles are updated at the beginning of every integration cycle. The 
macroparticles are propagated as 
ordinary tree particles while the chain subsystems are propagated after the 
drift operation, before updating accelerations of the active tree particles. The 
force 
corrections from the resolved macroparticles are computed after the acceleration 
calculation.

\subsection{Multiple chain subsystems}

As \ketju{}  allows for multiple chain systems in a single
simulation, it is possible for two chain subsystems to first
perturb each other and then eventually merge. The tidal 
perturbations from one chain subsystem on another are treated in the following 
way. We
resolve the macro particle into its constituent chain particles and treat them 
as described in \S \ref{perturbers}. Finally, we merge the chain subsystems 
labeled 
$i$ and $j$ into a single subsystem if the
volumes occupied by the chain systems overlap:
\begin{equation}
 \norm{\vect{r}_{\mathrm{\bullet},i} - \vect{r}_{\mathrm{\bullet},j}} < 
r_{\mathrm{chain},i} + r_{\mathrm{chain}, j}
\end{equation}
while the center-of-mass separation of the subsystems is decreasing, i.e. 
\begin{equation}
\left(\vect{r}_{\mathrm{\bullet},i} - \vect{r}_{\mathrm{\bullet},j} \right) 
\cdot 
\left(\vect{v}_{\mathrm{\bullet},i} - \vect{v}_{\mathrm{\bullet},j} \right) < 
0. 
\label{eq:approaching}
\end{equation}
We test for these conditions for all the chain subsystems at every 
\gadget{} timestep. Likewise, we split a chain subsystem into two new 
subsystems 
if
\begin{equation}
 \norm{\vect{r}_{\mathrm{\bullet}, i} - \vect{r}_{\mathrm{\bullet}, j}} >
r_{\mathrm{chain},
i} + r_{\mathrm{chain}, j}
\end{equation}
and the splitting SMBH $j$ is receding from all the SMBHs in the original 
subsystem, i.e. the condition
\begin{equation}
\left(\vect{r}_{\mathrm{\bullet},i} - \vect{r}_{\mathrm{\bullet},j} \right) 
\cdot 
\left(\vect{v}_{\mathrm{\bullet},i} - \vect{v}_{\mathrm{\bullet},j} \right) > 0
\label{eq:receding}
\end{equation}
must be fulfilled for every pair of SMBHs $i \neq j$.

\subsection{Particle mergers}

In the standard \gadget{} implementation \citep{springel2005}, the SPH kernel 
of 
the code is also used to compute the gas density around the SMBHs. In addition, 
the size $h$ of the kernel also defines the SMBH merging criterion. Two SMBHs 
merge if they come within a distance of $h$ of each other and the relative 
speed 
of the SMBHs is below the local sound speed. This typically occurs at SMBH 
separations of the order 
of tens or hundreds of parsecs (e.g. \citealt{Mayer2007, Johansson2009b}). 
Because the gravitational forces in the chain subsystems 
are not softened, we are able to follow the orbital evolution of a SMBH binary 
to separations well below the gravitational 
softening length $\epsilon$ of the tree calculation, whereas in a softened 
simulation the binary would 
stall at the softening length. Since we can also resolve arbitrarily close 
encounters between two SMBHs and between SMBHs and stellar particles, a more 
refined criterion for mergers between SMBHs and SMBHs and stellar particles is 
now required. 

If Post-Newtonian corrections of the order PN2.5 or higher are 
included, the semi-major axis 
of a point mass binary will shrink due to the loss of orbital energy caused by 
gravitational wave emission. 
The time evolution of the orbital semi-major axis, $a$, can be approximated by 
the 
analytical formula of 
\citet{peters1963} valid at PN2.5: 
\begin{equation}
    \dot{a} = - \frac{64}{5} \frac{G^3 M_{\bullet,1} M_{\bullet,2}
    (M_{\bullet,1}+M_{\bullet,2})}{c^5
a^3} \frac{1+\frac{73}{24} e^2 + \frac{37}{96} e^4}{(1-e^2)^{7/2}},
\label{eq: peters}
\end{equation}
where $M_{\bullet,1}$ and $M_{\bullet,2}$ are the masses of the two SMBHs
and $e$ is the eccentricity of the SMBH binary. 
Since $\dot{a}\propto -a^{-3}$, the coalescence timescale can be
approximated by
\begin{equation}\label{eq:tcapprox}
    t_c \sim -\frac{a}{4\dot{a}},
\end{equation}
if constant eccentricity is assumed.

For each bound SMBH binary, we compute the orbital elements and the
coalescence timescale using Eq. \eqref{eq:tcapprox}
before each global \gadget{} timestep. We compare the coalescence timescale 
$t_\mathrm{c}$ with the current tree
timestep $\dttree$ multiplied by a temporal safety factor $s_1>1$.
If $t_\mathrm{c} < s_1\dttree$, we merge the SMBHs instantly during this 
timestep.
The same coalescence criterion is applied to the stellar particles bound to a 
SMBH as well.
The safety factor is necessary, since Eq. \eqref{eq:tcapprox} only 
gives an approximation to
the coalescence timescale, and the exact dynamics might bring the
particles to a collision within $\dttree$ even though fiducially $t_\mathrm{c} >
\dttree$. We set $s_1=2$ in the code to ensure that this unphysical behavior
does not take place. For the simulations presented in this study, the expected 
absolute error in the SMBH merger timescale is conservatively a few times the 
length of a typical timestep $\sim$ 0.001 Myr. The typical SMBH merger 
separation in \ketju{} is of the order of a few hundred AU, which is three to 
four orders of magnitude below the typical merger separations in \gadget{} 
simulations. 

We further enforce a minimum distance between two particles. For a SMBH-SMBH 
pair, we use a multiple of 
the sum
of the Schwarzschild radii of the particles and set 
\begin{equation}    
    r_{\text{min,S}} = 6\left(\frac{2GM_{\bullet,1}}{c^2} +
    \frac{2GM_{\bullet,2}}{c^2}\right).
\end{equation}
This criterion is based on standalone tests with the \archain{} integrator,
which indicate that the Post-Newtonian dynamics is still numerically 
well-behaved at these distances.
For a pair consisting of a SMBH and a stellar particle, we use
\begin{equation}
    r_{\text{min,T}} = \max\left\{r_{\text{min,S}},\,\,
        s_2 \Rsun \left(\frac{M_\bullet}{\Msun}\right)^{1/3}
        \right\},
\end{equation}
where $s_2>1$ is a spatial safety factor and $\Msun$ and $\Rsun$ are the solar 
mass and 
radius, respectively. This criterion is motivated by the usual
definition for the stellar tidal disruption
radius, $r_\text{TDE}\sim (M_\bullet/m_\star)^{1/3} R_\star$,
assuming $R_\star=(m_\star/\Msun)^{1/3}\Rsun$ for the stellar
particles. For $s_2\sim 1$, the
criterion reduces exactly to the tidal disruption distance
\citep[e.g.][]{kesden2012}. To enforce larger separations well above the tidal 
disruption distance, we set $s_2=10$ in the code. This was 
motivated by the fact that the PN-corrections were found to be occasionally 
numerically unstable in the case of two-body collisions, combined with the fact 
that collisions are checked for before each
tree timestep using a linear prediction of the particle orbits.
This linear prediction gives the condition
\begin{equation}\label{eq:mergecond2}
    2 t\,\vect{r}_{12}\cdot \vect{v}_{12} + \vect{r}_{12}\cdot \vect{r}_{12}
    - r_\text{min}^2 = 0,
\end{equation}
where $\vect{r}_{12} = \vect{r}_2-\vect{r}_1$ and $\vect{v}_{12} =
\vect{v}_2-\vect{v}_1$ are the relative positions and velocities of the
particle pair. If Eq. \eqref{eq:mergecond2} has a solution with
$t\in[0,\Delta t_\text{tree}]$, we merge the particles instantly.

The actual merger of the particles is performed using the equations 
\begin{align}
    M &= M_1+M_2 \\
    \vect{r} &= (M_1\vect{r}_1 + M_2\vect{r}_2) / M \\
    \vect{v} &= (M_1\vect{v}_1 + M_2\vect{v}_2) / M \\
    \vect{L} &= \frac{M_1 M_2}{M} \vect{r}_{12} \times \vect{v}_{12} \\
    \vect{S} &= \vect{L} + \vect{S}_1 + \vect{S}_2.\label{eq:merger_spin}
\end{align}
This ensures the conservation of Newtonian linear momentum and angular momentum 
$|\vect{L}|$.
We note here that \ketju{} can follow the spin $(\vect{S})$ evolution of all 
stellar
and black hole particles. The spin state of the particles is only
affected by the PN corrections, through Eq. \eqref{eq:PN_spin}, and
for black holes also by the merger Eq. \eqref{eq:merger_spin}.
However, in the simulations run in this study, all particle spins are 
initialized
to zero. While the stellar spins remain zero at all times, the black hole spins 
also never attain a significant magnitude in the simulation.

\section{The regularized integrator}
\label{ARCHAIN_int}
\subsection{Algorithmic chain regularization}
\label{ARCHAIN}

The regularized dynamics in \ketju{} is based on a novel 
reimplementation of the \archain{} algorithm \citep{mikkola2008} written in
the C programming language. Below we will briefly outline the algorithm. For a 
more
involved description, see 
\citet{mikkola1993} and \citet{mikkola2006,mikkola2008}.
The algorithm has three main aspects:
algorithmic regularization, the use of relative distances to reduce
round-off errors, and extrapolation to obtain high numerical accuracy in 
orbit integrations. 

Algorithmic regularization works by transforming the equations of motion into a 
form
where integration by the common leapfrog method yields exact orbits to within
numerical precision for a Newtonian two-body problem, including
two-body collisions. This is achieved by introducing a new fictitious time as an
independent variable in order to circumvent the collision singularity that 
plagues the
Newtonian equations of motion. For mathematical details of the time 
transformation procedure, see Appendix~\ref{sc:algoreg}.

The second aspect of the regularization scheme is the use of relative
positions in the numerical calculations to reduce the often significant effect 
of
round-off error. The
relative positions, or chain vectors, form a contiguous `chain' of vectors,
\begin{equation}
    \vect{X}_k = \vect{r}_{j_k} - \vect{r}_{i_k},
\end{equation}
where $j_k$ and $i_k$ reindex the particles into endpoints and starting
points of the chain vectors, respectively. Here $k=1,\ldots,N_{\mathrm{c}}-1$, 
since there 
is 
one
fewer chain vector than there are particles. In the following, we assume
that the particles have been reordered so that $\vect{X}_k =
\vect{r}_{k+1}-\vect{r}_{k}$.
The formal Newtonian equations of motion for the chain variables are then
\begin{equation}
    \begin{aligned}
    \dot{\vect{X}}_i &= \vect{V}_i\\
    \dot{\vect{V}}_i &= \vect{A}_i(\{\vect{X}_i\}) + \vect{f}_i,
    \end{aligned}\label{eq:chaineqs}
\end{equation}
where $\vect{V}_i$ are the relative  velocities, $\vect{A}_i$ gives the
gravitational N-body acceleration from the chain particles and $\vect{f}_i$ 
incorporates all perturbing accelerations. These typically include
accelerations induced by simulation particles not contained within the
chain, since only a small subset $(\sim 10-100)$ of all simulation particles
are found in the chain at any given time. 

Up to this point, the result is just a reformulation of
the original problem. The defining aspect is how the chain vectors
are chosen. The selection criteria can be based on either the relative
distances or the magnitudes of the forces between the particles, so that
the shortest distances or the strongest forces are included in the chain, 
respectively.
If relative distances are used as the selection criterion, the algorithm 
proceeds as follows:
\begin{enumerate}
    \item Find the shortest relative distance between two particles in a 
subsystem.
        This forms the first segment of the chain, where the two
        particles are now called the `head' and the `tail' of the chain.
    \item From the subsystem particles not yet in the chain, find the particle
        with the shortest relative distance to the head (tail)
        particle, and add it to the chain. This particle is now the new
        head (tail).
    \item Repeat step 2 until no more particles remain.
\end{enumerate}
If instead forces between the particles are used, the algorithm is exactly the
same, except ``shortest distance'' is replaced by ``strongest 
force''.
The new variables $\vect{X}_i$ are then propagated using Eqs. 
\eqref{eq:chaineqs}. 
They are also used in place of the $\vect{r}_{ij}$ in all
calculations where relative distances are required, as long as
at most $N_{\mathrm{d}}$ chained distances $\vect{X}_i$ need to be added to 
yield
$\vect{r}_{ij}$.
The actual summation can be done using the following prescription:
\begin{equation}\label{eq:chaindist}
    \vect{r}_{ij} =
    \left\{\begin{aligned}
            &\vect{r}_j - \vect{r}_i & \text{if $\abs{i-j} > N_{\mathrm{d}}$} \\
            &\sum_{k=\min\{i,j\}}^{\max\{i,j\}-1}\hspace{-3ex}
\sgn(i-j)\vect{X}_k & \text{if $\abs{i-j} \leq N_{\mathrm{d}}$}.
\end{aligned}\right.
\end{equation}
Following \citet{mikkola2008}, we set $N_{\mathrm{d}}=2$.
The regularized leapfrog algorithm using the chained
variables is described in detail in Appendix~\ref{sc:chain_leapfrog}.

The final ingredient in the algorithmic chain regularization is
the use of an extrapolation method. In broad terms, this entails taking
a longer timestep $H$ and subdividing it into $n$ smaller steps, each of
which is performed using some suitable numerical method, such as
the modified midpoint method.
When the subdivision count $n$ is successively increased, the result
will generally converge towards the exact solution of the equations of
motion over the longer timestep $H$. The results of this process can then be 
extrapolated to
$n\fromto\infty$ using either rational or polynomial extrapolation. This
method is called the Gragg--Bulirsch--Stoer (GBS) algorithm
\citep{gragg1965,bulirsch1966}.
Practical implementations include
sophisticated timestep control as well as some control of the maximum
subdivision count, which turns out to be proportional to the order of the
method \citep[see e.g.][]{hairer2008}.
For a thorough exposition of the GBS extrapolation method, as well as a
complete implementation see \citet{press2007}. We use this 
implementation in our code with the added modification of using the 
chain leapfrog (see Appendix~\ref{sc:chain_leapfrog}) instead of
the modified midpoint method to propagate the system through the
substeps.

When combined, the three aspects of the chain regularization method
guarantee that two-body collisions are treated exactly up to numerical
precision, round-off errors are greatly reduced and the desired tolerance
for energy errors during the propagation can be set to a very low level without
excessive degradation of the performance of the algorithm.

\subsection{Post--Newtonian corrections}\label{sc:PN}

In \ketju, we implement relativistic
corrections to motions near black hole particles via the so-called
Post--Newtonian corrections. These are represented by additional terms in
the relative acceleration of two bodies, approximating the effects of general
relativity, so that
\begin{equation}
    \vect{a}_\text{2-body} = \vect{a}_\text{Newtonian} + \sum_{k=2}^{7}
    c^{-k}\vect{a}_{(k/2)\text{PN}} + \vect{a}_S,
\label{PN_terms}
\end{equation}
where $\vect{a}_\text{Newtonian}$ is the usual Newtonian two-body
acceleration, $c$ is the speed of light, $\vect{a}_{x\text{PN}}$ is the
PN correction of order $x$ and $\vect{a}_S$ indicates PN terms depending
on the spins of the particles.
We include both spin-independent and spin-dependent PN corrections up to
order PN3.5 corresponding to inverse seventh power of the speed of light, i.e.
$c^{-7}$ (see e.g. \citealt{Will2006} for further details).

In addition, for spinning bodies, there is a corresponding PN
contribution to the equations of motion for the spins, given by
\begin{equation}
\dot{\vect{S}}_i = \vect{S}_{\text{PN},i} \times \vect{S}_i,
\label{eq:PN_spin}
\end{equation}
where $\vect{S}_i$ is the spin angular momentum of the particle $i$
and $\vect{S}_{\text{PN},i}$ gives the effect of the spin--orbit,
spin--spin and quadrupole--monopole interactions.
The explicit forms for the included PN terms can be found
in \citet{mora2004} for the spin-independent terms and
\citet{barker1975} and \citet{kidder1995} for the spin-dependent terms.
The two-body PN corrections in Eq.~\eqref{PN_terms} are only used for 
interactions where at
least one of the bodies is a black hole particle. For interactions
between stellar, gas or dark matter particles, the PN corrections are
not expected to be of any significance, at least in the physical scenarios
for which the \ketju{} code is intended.

The code also provides the option of using the PN cross-term
formulation \citep{will2014} instead of the two-body formulation given above. 
The cross terms are an approximation of the full Einstein--Infeld--Hoffman 
(EIH) 
equations of motion \citep{einstein1938} and 
are valid at PN2.0 order. In addition, the approximation is only 
valid for a system consisting of one or a few very massive bodies and numerous 
lighter bodies. As such, it is in particular suitable for systems consisting of
SMBHs surrounded by lighter stellar particles. 
In practice, the cross terms modify the Newtonian two-body acceleration
between two particles by the first order PN terms, as well as an
additional contribution depending on the accelerations of all the other
particles in the subsystem.
Similarly to the EIH equations, the cross terms involve
sums with $\bigO(N_{\mathrm{c}}^3)$ terms for a system of $N_{\mathrm{c}}$ 
particles,
albeit with a smaller proportionality constant. The cross term
contributions can be used only for a modest number of particles, of
the order of hundreds at most, without prohibitive loss of numerical 
performance. 

\section{Test problems and code calibration}
\label{Test_code}

We calibrate our user-specified \ketju{} parameters which control the chain 
radius, the perturber radius and the \archain{} integrator accuracy for our 
regularized 
tree code by comparing our code against the standard gravitational collisional 
N-body simulation code NBODY7 \citep{aarseth2012}. This is a gravitational 
direct summation code utilizing an accurate 
fourth-order Hermite integrator with force polynomials and few-body 
regularization for 
close encounters of simulation particles. The employed few-body regularization 
method 
is optionally either the algorithmic chain or the Kustaanheimo-Stiefel (KS) 
regularization method \citep{kustaanheimo1965}. The current publicly available 
version of 
NBODY7 is accelerated with the Ahmad-Cohen neighbor scheme and GPUs 
\citep{aarseth2012}.
In addition, for comparison, we also run tests with the standard version of 
\gadget{} \citep{springel2005b}
without including the chain regularization. The test and calibration setups 
used 
in this section closely follow 
the performance tests presented by \cite{karl2015}, 
which were used to verify the performance of the regularized tree code rVINE.

\subsection{The inspiral of a single SMBH in a Hernquist sphere}
\label{inspiral_Hernquist}
We consider first a SMBH on a circular orbit in a Hernquist sphere 
\citep{hernquist1990}. 
A SMBH propagating in a field of stars is subject to 
dynamical friction (\citealt{chandrasekhar1943}, \citealt{binney2008}) and will 
sink to the center of the Hernquist bulge on the dynamical friction timescale. 
Throughout this section we use the following Hernquist model for our 
calibration tests:
total mass of $M = 10^{10}$ $M_{\odot}$ and a scale radius of 
$a=1$ kpc. A multi-component extension of the single-component  Hernquist 
profile is discussed in \S 6.
In this Hernquist sphere we place a single SMBH with 
a mass of $M_{\bullet}=10^7$ $M_{\odot}$ initially on a circular orbit 
($v_{\mathrm{circ}} \approx 95.4$ km/s) at the half-mass 
radius ($r_{\mathrm{H}} = \left(1+\sqrt{2} \right) a \approx 2.41$ 
kpc) of the Hernquist sphere. 

\begin{figure}[h!]
\includegraphics[width=\linewidth]{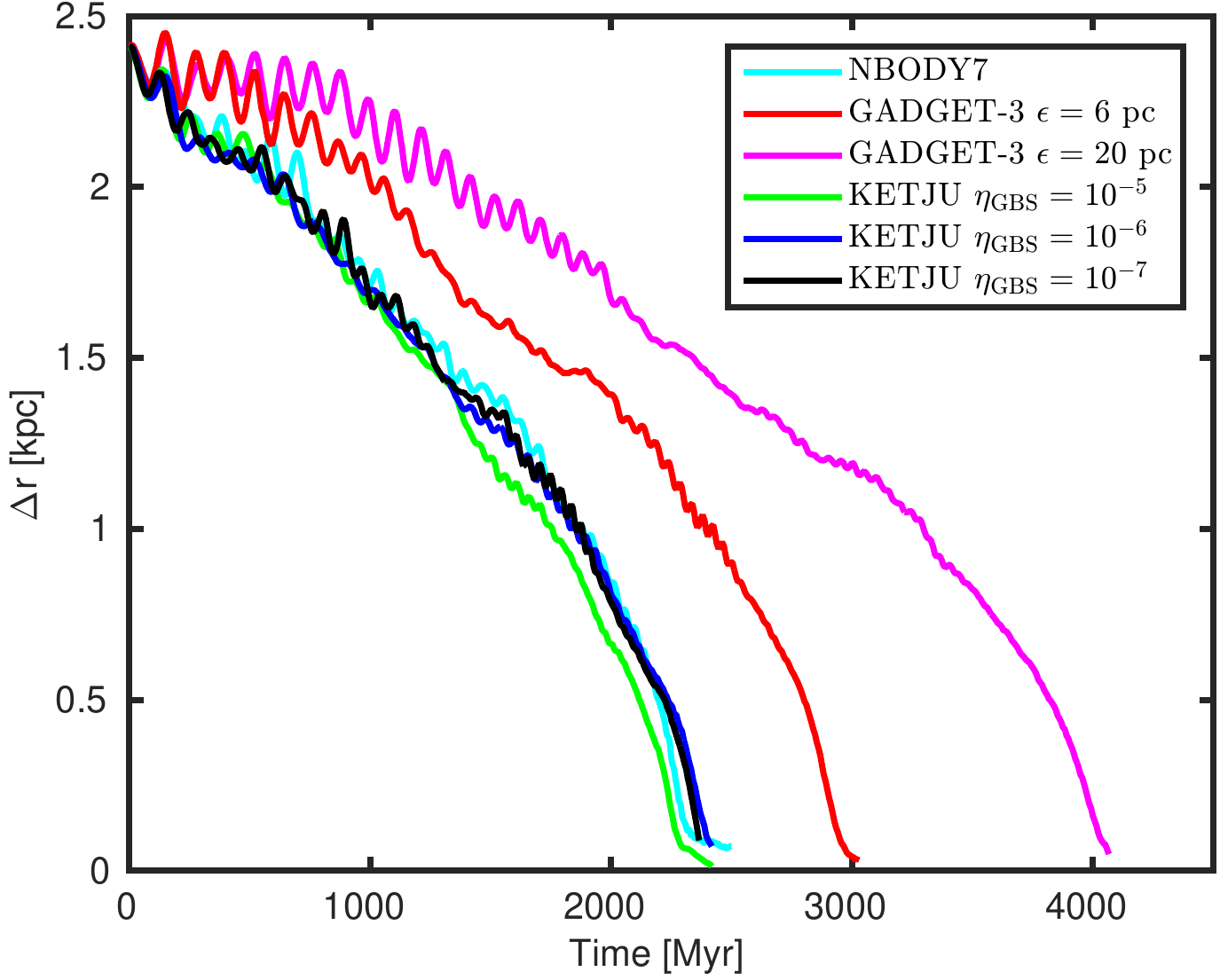}
\caption{The separation from the center of the galaxy as a function of 
simulation 
time
demonstrating the sinking of a SMBH towards the center of a Hernquist sphere 
using three different numerical codes: NBODY7, \gadget{} and the \ketju. 
All three 
displayed \ketju{} simulation runs with different GBS integrator accuracies 
match the SMBH 
sinking time scale in NBODY7 within a few percent, whereas the \gadget{} run 
overpredicts the decay time by a factor of 1.25 to 1.7, depending on 
the chosen gravitational softening length.}
\label{fig: friction}
\end{figure}

The number of particles in the dynamical friction test setup is restricted to 
$N = 10^5$ because of the steep scaling of the required computational time in 
NBODY7 as a function of the particle number. We run the dynamical friction 
calibration simulations using NBODY7, standard \gadget{} and \ketju{} until the 
SMBH reaches the center of the Hernquist sphere 
where the 
dynamical friction becomes ineffective. The results of the dynamical friction 
runs are presented 
in Fig.~\ref{fig: friction}. Throughout these tests the NBODY7 run acts as the 
standard 
against which other codes are compared.

For the \gadget{} runs, we test two different gravitational softening 
lengths: $\epsilon= 6$ pc and $\epsilon = 20$ pc. We set the \gadget{} 
integrator error tolerance to $\eta = 0.02$ and the force accuracy to $\alpha = 
0.005$, using the standard \gadget{} cell opening criterion 
\citep{springel2005b}, in all dynamical friction runs. We also run several 
simulations with higher values for the accuracy parameters obtaining 
consistently similar results as with the 
standard parameter values. In the \gadget{} run with $\epsilon = 20$ pc 
the SMBH reaches the center of 
the Hernquist sphere on a sinking timescale of $t_{\mathrm{sink}} \sim 4000$ 
Myr, 
whereas for the run with $\epsilon = 6$ pc the sinking time is 
$t_{\mathrm{sink}} \sim 3000$ Myr.
For both runs the sinking timescales are considerably longer than the sinking 
timescale in the 
NBODY7 simulation ($t_{\mathrm{sink}} \sim 2400$ Myr).
This is due to the fact that the dynamical friction force is weaker 
for softened gravitational interactions than for the non-softened forces 
\citep{just2011}, as in NBODY7. In simulation codes using gravitational 
softening, the 
dynamical 
friction force contribution of the stars with an impact parameter smaller than 
the gravitational softening length is grossly underestimated. This results in 
reduced dynamical friction and affects the sinking timescale of the SMBH 
although the friction force depends only logarithmically through the Coulomb 
factor on
the impact parameter of the encounters of the SMBH and stellar particles 
\citep{binney2008}.

Including a regularized region around the SMBH overcomes the limitations 
of the softened tree codes in the computation of the dynamical friction force. 
In \ketju{}, the far-field gravitational dynamics of \gadget{} remains 
unaltered while the regularized \archain{} integrator handles the close 
encounters between the SMBH and the incoming stars. We set the 
gravitational softening length to $\epsilon = 6$ pc and the chain radius of 
the SMBH to be constant at $r_{\mathrm{chain}} = 30$ pc 
($\lambda = 30$, see Eq. \eqref{rinfl}). The two important numerical 
parameters that need to be calibrated are the tidal 
parameter $\gamma$ and the Gragg-Bulirsch-Stoer (GBS) extrapolation accuracy 
parameter \etabs (see \S \ref{ARCHAIN}). The tidal parameter $\gamma$ 
defines the size of the 
perturber volume around the regularized subsystem according to Eq. 
\eqref{tidalgamma}. The GBS accuracy parameter $\etabs$ sets the 
maximum 
allowed error during a single \archain{} step for any physical variable
\citep[see][for an in-depth description of the GBS accuracy
parameter]{press2007}.

We run the dynamical friction test using \ketju{} with three different GBS 
accuracy parameters \etabs{} $\in [10^{-5}, 10^{-6}, 10^{-7}]$. We set the 
tidal parameter $\gamma$ so that the perturber radius equals twice the radius 
of the chain subsystem, i.e. 60 pc, which yields good results for all 
the 
dynamical 
friction test runs. During the 
first $\sim 1.5$ Gyr, the SMBH propagates through the low-density outer parts 
of the Hernquist sphere and the chain regularization is needed only 
occasionally when a stellar particle passes very close to the sinking SMBH. 
After $t > 2$ Gyr the regularized subsystem contains particles at every global 
\gadget{} timestep. We obtain final SMBH sinking times that are 
within $4\%$ of the NBODY7 result of $t_{\mathrm{sink}}=2.4$ Gyr using GBS 
parameter values of \etabs{} $\leq 10^{-5}$.

\begin{figure*}
\includegraphics[width=\textwidth]{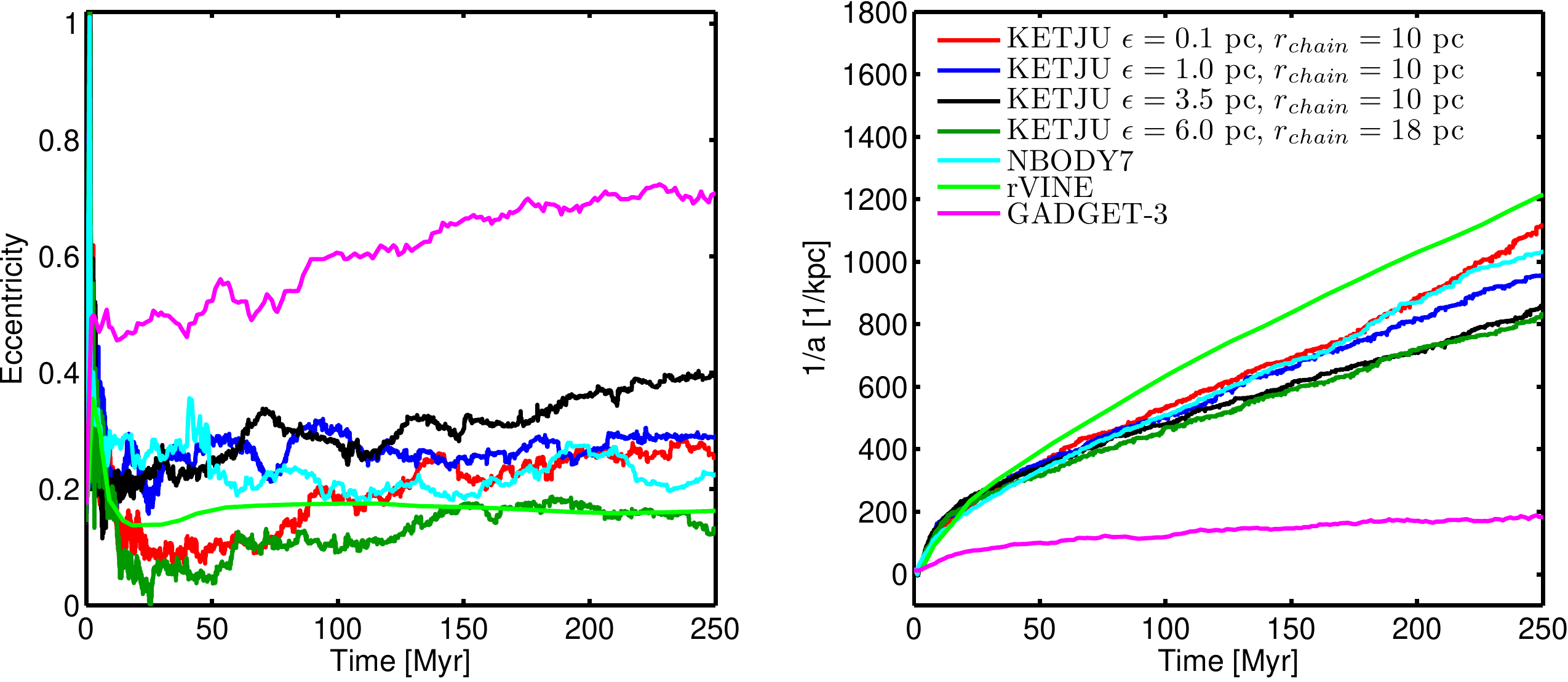}
\caption{The evolution of a SMBH binary simulated using NBODY7, 
standard \gadget{}, rVINE and four \ketju{} runs. In all \ketju{} simulations
the GBS error tolerance is set to \etabs $=10^{-6}$. 
Left panel: the binary 
eccentricities are in general small, except for the \gadget{} run. Right panel: 
the evolution of the inverse semi-major axis. When the gravitational softening 
is very small ($\epsilon \lesssim 1$ pc), the \ketju{} result is close to 
the NBODY7 result. When $\epsilon$ is increased to $3-5$ pc, the \ketju{} 
results appear to converge to a slightly gentler hardening slope than seen in 
the NBODY7 run. Runs with \ketju{} and NBODY7 for a low resolution Hernquist 
sphere with $N=10^5$ particles results in unphysically 
strong two-body relaxationand steep hardening slopes, when $\epsilon 
\rightarrow 
0$. In the \gadget{} run, the SMBH binary stalls around $1/a \sim 1/\epsilon$, 
as 
expected.
}
\label{fig: hardening}
\end{figure*}

\begin{figure*}
\includegraphics[width=\textwidth]{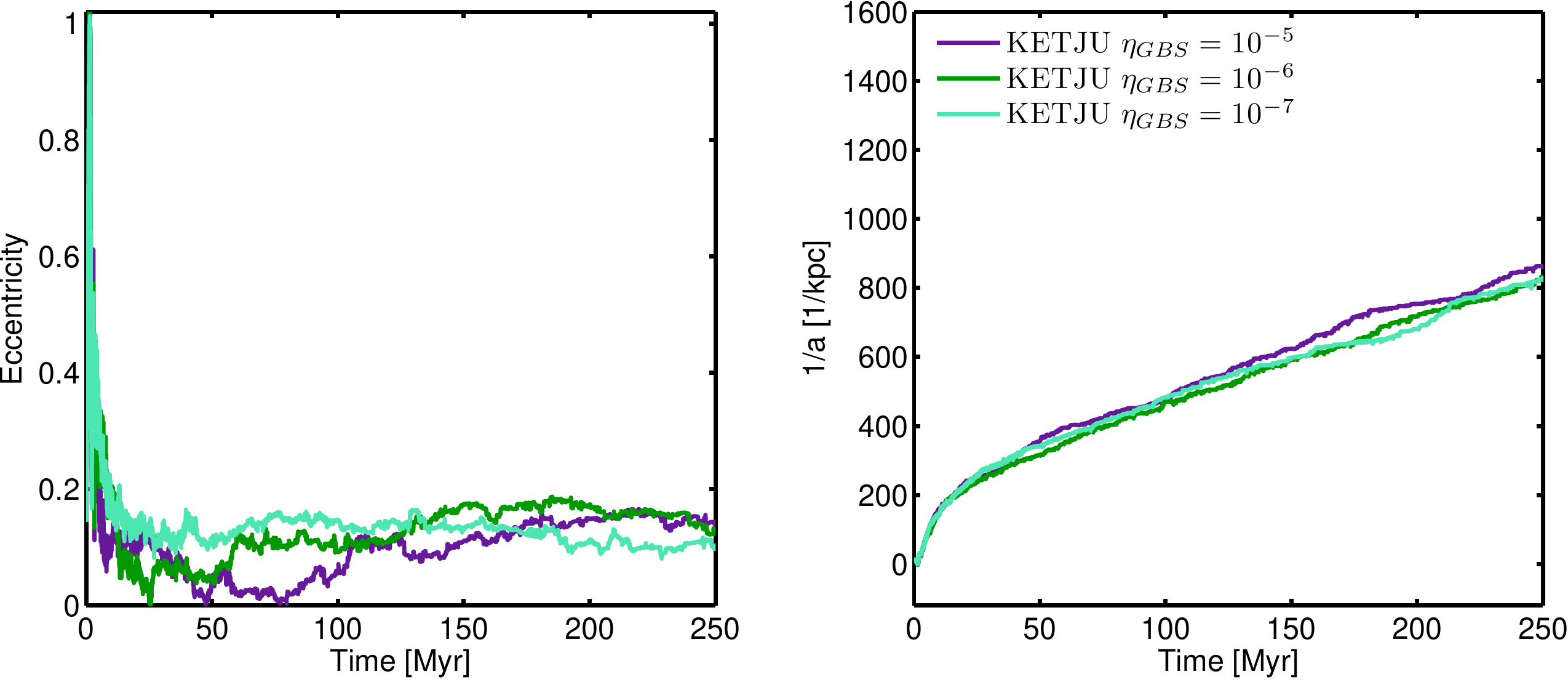}
\caption{The effect of the \ketju{} GBS error tolerance parameter on the SMBH 
binary 
evolution. The gravitational softening length is fixed to $\epsilon = 6$ pc and 
the chain radius is set to $r_{\mathrm{chain}} = 18$ pc. The run with \etabs 
$=10^{-6}$ is the same as in Fig. \ref{fig: hardening}. All \ketju{} runs with 
different
\etabs{} values, \etabs $=10^{-5}$, \etabs $=10^{-6}$ and \etabs $=10^{-7}$, 
yield 
consistently 
similar results both in the evolution of binary eccentricity and the semi-major 
axis.
}
\label{fig: ecc}
\end{figure*}

\subsection{A SMBH binary hardening in a Hernquist sphere}

Another crucial feature for a regularized tree code is 
the ability to properly model the formation and the hardening of systems of 
binary (or multiple) SMBHs. We build the initial conditions for a SMBH binary 
hardening test using the same Hernquist spheres as in the previous section. 
We note that the simulation particle number $N=10^5$, 
limited by the scalability of NBODY7, might be too low to properly study the 
SMBH binary evolution. Recent state-of-the-art direct summation 
studies (e.g. \citealt{khan2013, Vasiliev2014b}) utilizing $\phi$GRAPE-based 
codes \citep{Harfst2007, Harfst2008} have employed particle numbers up to $N 
\sim 10^6$, but as demonstrated by \cite{Vasiliev2015} and 
\cite{gualandris2016} even simulations with these high particle numbers are
affected by spurious relaxation effects. Instead, our main goal here is to 
demonstrate that \ketju{} can reproduce NBODY7 results in simulation setups 
which are possible to run using NBODY7 in a reasonable wallclock time.

One SMBH with mass of $M_{\bullet} = 5 \times 10^7$ $M_{\odot}$ is placed at 
rest at the center of 
the Hernquist sphere while another SMBH of the same mass is placed on 
a circular orbit with an initial separation of $r = 0.1$ kpc from the center of 
the sphere. We run the 
simulation for $t = 250$ Myr after which the separation of the two SMBHs is 
$\sim 1$ pc. In the simulation runs using \gadget{} we set the gravitational 
softening length of the SMBHs and stellar particles to $\epsilon = 10$ pc.
The parameter study of \ketju{} is twofold. First, we set the GBS tolerance to 
\etabs $=10^{-6}$ and test the effect of the gravitational softening length and 
the chain radius on the SMBH binary evolution. We try four different softening 
lengths: $\epsilon = 0.1$ pc, $\epsilon = 1.0$ pc, $\epsilon = 3.5$ pc and 
$\epsilon = 6.0$ pc. The chain radius 
is fixed at $r_{\mathrm{chain}} = 10$ pc in the former three runs and set to 
$r_{\mathrm{chain}} = 18$ pc in the last run with the largest softening length. 
The perturber radius is 
twice the chain radius in all the simulation runs. The results of this test are 
presented in Fig. \ref{fig: hardening}.

In the \gadget{} run the SMBH binary stalls at a separation of $\sim \epsilon$, 
as expected. In 
addition, the binary eccentricity is higher in the \gadget{} run when compared 
to 
the rVINE, NBODY7 and \ketju{} simulations. In rVINE the evolution of the 
inverse 
semi-major axis depends on the initial chain radius \citep{karl2015}. With 
$N=10^5$ stellar particles, the final SMBH binary inverse semi-major axis is 
somewhat larger in the rVINE run than in the NBODY7 simulation run. 
As expected, the \ketju{} runs with the smallest softening lengths match best 
the evolution of the inverse semi-major axis in NBODY7. With the two larger 
softening lengths ($\epsilon = 3.5$ pc and $\epsilon = 6$ pc) the hardening 
rate appears to converge to a slightly lower value than in the NBODY7 run.

As the host galaxy is a low-resolution Hernquist sphere, the dominating 
loss-cone 
filling effect is two-body relaxation. Increasing the gravitational 
softening 
length reduces the loss-cone filling rate by increasing the two-body 
relaxation timescale. Thus it is natural that the hardening rate decreases when 
the 
softening length is increased. In a typical real spherical galaxy the two-body 
relaxation timescale is very long because the number of stars is $N \gg 10^5$ 
and 
thus the resulting loss-cone filling would be very inefficient, with the 
hardening rate 
going towards zero as $N$ increases. However, typically real SMBH binaries form 
in the aftermaths 
of galaxy mergers, where the non-spherical shape of the host galaxy is the 
primary driver for the
loss-cone filling instead of two-body relaxation (e.g. \citealt{khan2011}). 
Thus, we here
argue that the small differences between the \ketju{} hardening rates and the 
NBODY7 results are not a 
problem when simulating more physical SMBH binary formation scenarios, such as 
the major 
mergers of galaxies.

The second part of the \ketju{} SMBH binary parameter study consists of varying 
the GBS error tolerance parameter. The gravitational softening was set to 
$\epsilon = 6$ pc and the chain radius to $r_{\mathrm{chain}} = 18$ pc for 
these 
runs. We 
tested three different tolerance parameter values: \etabs $=10^{-5}$, \etabs 
$=10^{-6}$ and 
\etabs $=10^{-7}$. The results presented in Fig. \ref{fig: ecc} show that  
the evolution of both the binary eccentricities and the inverse of the 
semi-major axis are quite similar for the
three runs. In general the orbital eccentricities of the SMBH binaries are 
quite 
low $(e<0.4)$ in all the test
runs. We do not see any apparent convergence of the results. However, this is 
not unexpected, as the eccentricity evolution 
of the SMBH binary is strongly dependent on the velocity distribution of the 
stellar 
component \citep{mikkola1992}, with the large scatter in the eccentricity just 
highlighting the low mass resolution of this set of simulations.

In conclusion, all the tested \ketju{} parameter 
combinations provide a significantly better description of the SMBH binary 
dynamics 
than standard \gadget{}, for which the SMBH binary separation is constrained by 
the 
gravitational softening length. \ketju{} also accurately reproduces the results 
of NBODY7. Based on these test simulations, we choose the GBS accuracy 
parameter 
of \etabs $= 10^{-6}$ for the rest of the \ketju{} simulations in this study.

\section{Performance and scalability}
\label{performance}

\subsection{Conservation of energy}

\begin{figure}[h!]
\includegraphics[width=\linewidth]{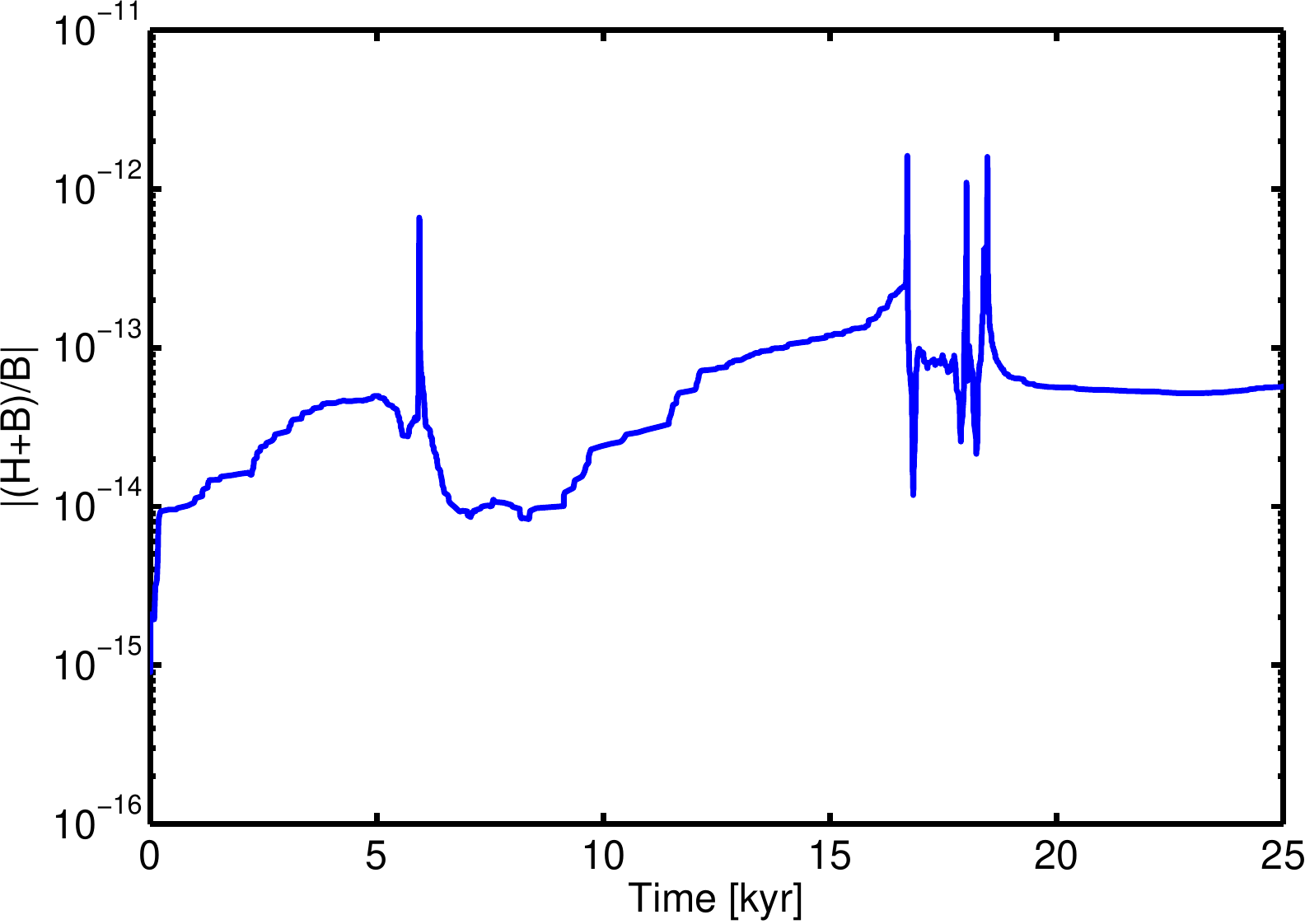}
\caption{The relative error of our \archain{} implementation in the extreme 
accuracy test of \citet{mikkola2008}. Here $B$ is the binding energy of the 
system and $H$ is the Hamiltonian. The relative energy error remains below 
$10^{-13}$ for most of the simulation time, excluding the 
most close-by encounters, which are clearly seen as visible jumps in the
energy error.}
\label{fig: energy_chain}
\end{figure}

The earlier FORTRAN-based implementation of the standalone \archain{} algorithm 
has been demonstrated to conserve energy extremely accurately in few-body 
($N_{\mathrm{c}}<10$) simulations \citep{mikkola2008}. Here we adopt the same 
test scenario 
for our C-based \archain{} reimplementation. We set a single SMBH with 
$M_\bullet = 10^{10} M_\odot$ at rest at the origin. 
Next, seven stellar particles logarithmically evenly spaced in their mass ratio
are drawn from the mass ratio range of $10^{-9} \le m_\star / M_\bullet \le 
10^{-3}$ 
and are placed around the SMBH with a zero initial velocity, resulting in 
almost rectilinear stellar orbits. We follow the dynamical evolution of the 
system for 
25 000 years. The energy conservation of the system is presented in Fig. 
\ref{fig: energy_chain}. The relative energy error remains $|\Delta E / E| = 
|(H+B)/B| \lesssim 
10^{-13}$ for most of the simulation time, where $H$ is the Hamiltonian, which 
corresponds to the negative of the gravitational binding energy $-B$ in the 
\archain{} implementation within numerical precision (see 
Appendix~\ref{sc:algoreg}.).

Conservation of energy has to be carefully ensured in \ketju{} as the 
gravitational potential in the simulation volume contains both softened and 
non-softened regions. The energy conservation in the standard softened TreePM 
algorithm of  \gadget \ was validated by \citet{springel2005b}. A simulation 
particle crossing the boundary from a softened potential to a non-softened 
regularized 
region (or vice versa) experiences a sudden discontinuity in the gravitational 
potential. We next demonstrate that the flux of simulation particles through a 
chain 
subsystem does not introduce additional error to the global energy conservation 
of the simulation. 

\begin{figure}[h!]
\includegraphics[width=\linewidth]{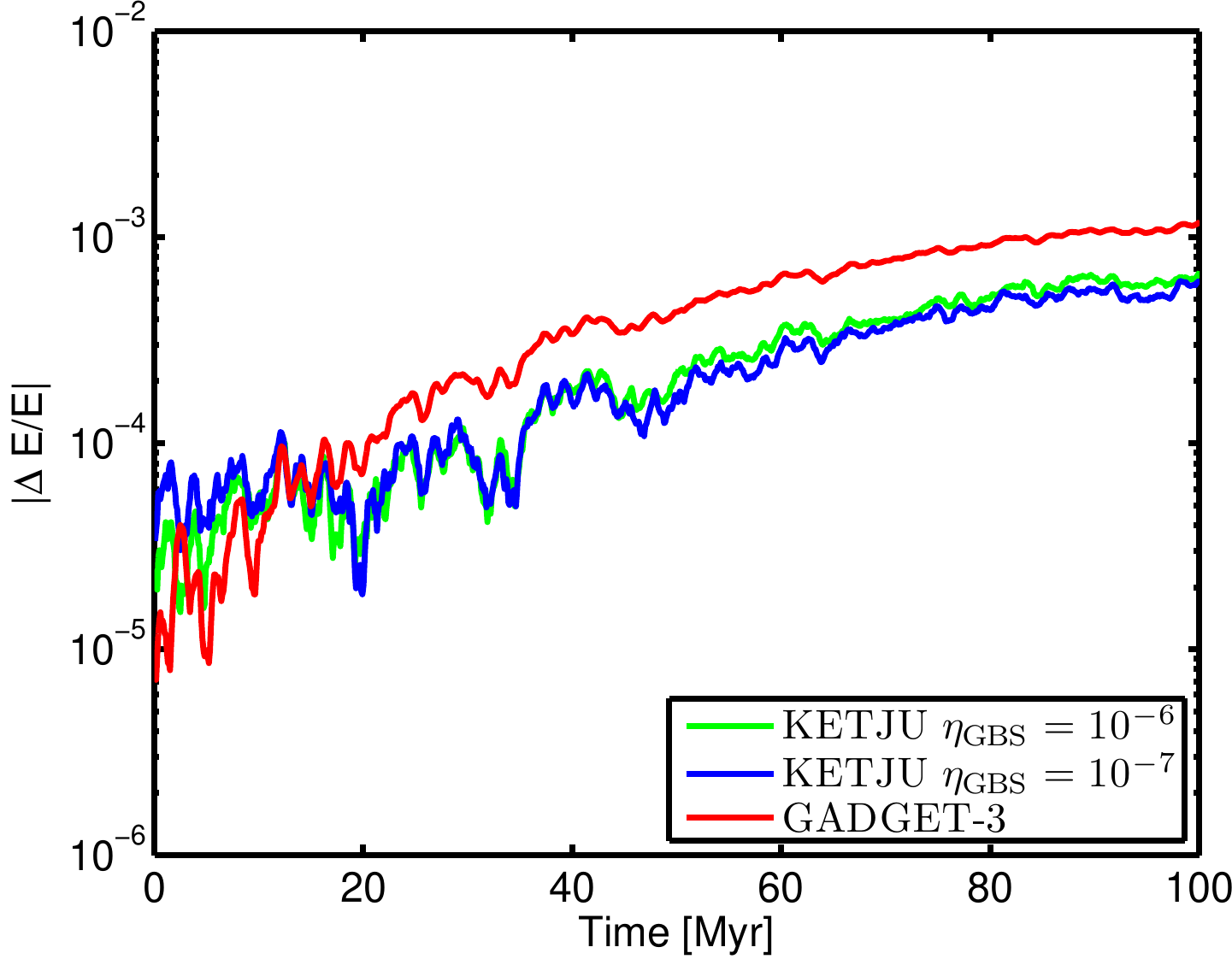}
\caption{The relative energy error in \gadget{} and \ketju{} in a 100 
Myr 
simulation of an isolated stellar bulge with a single SMBH. The energy 
conservation
is slightly better in all the \ketju{} runs, when compared to standard 
\gadget{}.}
\label{fig: energy}
\end{figure}

We set up an energy conservation test by constructing a Hernquist sphere as 
described in \S \ref{results} with $M = 10^{11} M_\odot$ and $N = 
10^6$ particles centered at the origin. A SMBH with $M_{\bullet} = 10^8 \Msun$ 
is placed at rest at the center of the sphere.  The gravitational softening 
length of stellar particles is set to $\epsilon = 6$ pc, the chain radius 
to 18 pc and the perturber radius to twice the chain radius, i.e. 36 pc.
We run the initial conditions for 100 Myr and study the energy conservation as 
a function of the GBS accuracy parameter. The relative error of the total 
energy 
$\abs{\Delta E/E_{\mathrm{0}}}$ as a function of 
time is presented in Fig. \ref{fig: energy}.

We find that energy is conserved in standard \gadget{} at the level of 
$\abs{\Delta E/E_{\mathrm{0}}} \lesssim 1\times 10^{-3}$ during the simulation. 
\ketju{} performs sligthly better: $\abs{\Delta E/E_{\mathrm{0}}} \lesssim 
7\times 10^{-4}$ with both \etabs $=10^{-6}$ and \etabs $=10^{-7}$. The 
difference between \gadget{} and \ketju{} clearly originates from the central 
tens of parsecs of the galaxy where the accelerations of the simulation 
particles are the strongest. As \ketju{} uses the GBS extrapolation method in 
\archain, the energy conservation up to a user-given tolerance is guaranteed in 
the regularized region during every timestep. In contrast, this is not the case 
with \gadget's 
leapfrog integrator: even though the timestepping is adaptive, there is no 
set maximum allowed energy error per timestep.

Our results can also be compared to the energy conservation of the rVINE code, 
for which 
\cite{karl2015} obtained an energy 
conservation of $5\times 10^{-4} \lesssim \abs{\Delta E/E_{\mathrm{0}}}
\lesssim 5 \times 10^{-3}$ with an initial chain 
radius of 10 pc in short test runs with a duration of 4.7 Myr and $N=10^5$ 
particles. The exact result depends on the chosen rVINE tree accuracy 
parameter, 
but the energy conservation values given above are representative. In both 
the \ketju{} test runs shown in Fig. \ref{fig: energy}, the energy is conserved 
at 
a level below 
$\abs{\Delta E/E_{\mathrm{0}}} \lesssim 8\times 10^{-5}$ during the first 5 Myr 
of the simulation. Based on our energy 
conservation tests, we conclude that \ketju{} conserves energy on a slightly 
better level than both standard \gadget{} and the rVINE code.

\subsection{Timing tests and code scalability}

In this section we demonstrate the scalability of the \ketju{} code for 
realistic simulation setups. The performance and scalability of collisionless 
standard \gadget{} simulations is presented in \cite{springel2005}. The most 
time-consuming operations of the \gadget{} code in simulations without gas are 
the computation of the gravitational force using the tree algorithm as well as 
the domain decomposition required for efficient simulation parallelization. 

\ketju{} introduces new computational tasks that need to be performed in 
addition to the standard \gadget{} procedures. The most important tasks from 
the perspective of CPU time consumption are the following: first, the 
gravitational 
oct-tree has to be built during every smallest timestep. This is 
necessary as 
the chain 
structure needs to be updated every timestep due to the possibilities of 
absorption of 
particles into the chain and the escape of particles from the chain. In 
addition, the required neighbor searches for chain particles and perturber 
particles and the resulting extra MPI (Message Passing Interface) communication 
typically consumes of the order of a few percent of the total CPU time. 

Our implementation of the \archain{} algorithm is MPI-parallelized for 
increased performance and compability with \gadget. The \ketju{} functions 
are implemented in \gadget{} in the same manner as all the other subresolution 
procedures: 
all MPI tasks participate in a single subresolution routine at a time. The order 
of the  
most important \archain{} function calls in the integration cycle of \gadget{} 
is briefly discussed at the end of 
\S \ref{timestepping_with_chain}. Every task contains a copy 
of the chain structure enabling fast chain-tree particle exhange calculations.
Each computational node performs the parallelized chain integration. 
This computation strategy is found to be faster than the parallelized chain 
integration using all the available tasks, or serial chain integration and 
communication of the results to all the tasks. Early development versions of 
\ketju{} used all the available MPI tasks for the chain integration, but this 
proved to be an 
extremely poor computational strategy when the number of chain particles was 
far below the number of MPI tasks.

The \archain{} integration of chain particles is the computationally
most demanding new operation introduced in \ketju{} compared to \gadget. 
Estimating the scaling of the 
computational demand with increasing chain and perturber particle numbers
$N_{\mathrm{c}}$ and $N_{\mathrm{p}}$ is not straightforward, since the 
\archain{} integrator
controls both the timestep and the order of the method as necessary to
stay within the set accuracy parameters. However, an estimate can be formulated 
as follows. The amount of required computational 
work 
per one force
calculation is of the order $\bigO(N_{\mathrm{c}}^2 + N_{\mathrm{c}} 
N_\mathrm{p})=\bigO(\tilde{N}^2)$, where
$\tilde{N} = \sqrt{N_{\mathrm{c}}^2 + N_{\mathrm{c}} N_\mathrm{p} }$ is now an 
effective 
number of chain particles.
As one force calculation is needed per timestep for a second order leapfrog, the
first estimate for the asymptotic computational scaling is just 
$\bigO(\tilde{N}^2)$.

Setting a tolerance limit on the error does not change this in the
first approximation. The error $\Delta x$ over one step in some
dynamical variable $x$ for a method of order $p$ is $\Delta x \propto h^{p+1}$,
where $h$ is the timestep. As such, the global error over a run of time
$T$ is $E_x \propto T h^{-1} \Delta x \propto h^p$. If we set an error
tolerance $\epsilon$ and demand $E_x \lesssim \epsilon$, we find that
$E_x\propto T^{1-1/p}\epsilon^{1/p}$ which is independent of the
particle number and gives a scaling $\bigO(\tilde{N}^2)$ again, if $p$ is
constant.

However, the force computation for each particle also suffers from
errors accumulated for all the other particles, leading to a force error
$\Delta F\propto \tilde{N}^2$, which then gives a total error for $x$
of $E_x \propto \tilde{N}^2h^p$. If we then demand $E_x \lesssim \epsilon$,
we find that $h \propto \epsilon^{1/p}\tilde{N}^{-2/p}$, resulting in a total
computational effort of $\propto T\epsilon^{-1/p}\tilde{N}^{2+2/p}$. 
Thus, the \archain{} algorithm scales as $\bigO(\tilde{N}^{2+2/p})$, where 
$p$ is now some mean order used by the GBS extrapolation scheme during the
run and which will depend on the smoothness of the problem. In general we have 
$p\geq 2$, and in the worst case scenario of a simple leapfrog, $p=2$, we have 
$\bigO(\tilde{N}^3)$
scaling. Typically the GBS method works at $p=10$ to $p=16$ which give
approximately $\bigO(\tilde{N}^{2.20})$ and $\bigO(\tilde{N}^{2.13})$, 
respectively. As such, we can estimate that the \archain{} integration should 
scale
approximately slightly worse than the square of the particle number.

\begin{figure}[h]
\includegraphics[width=\linewidth]{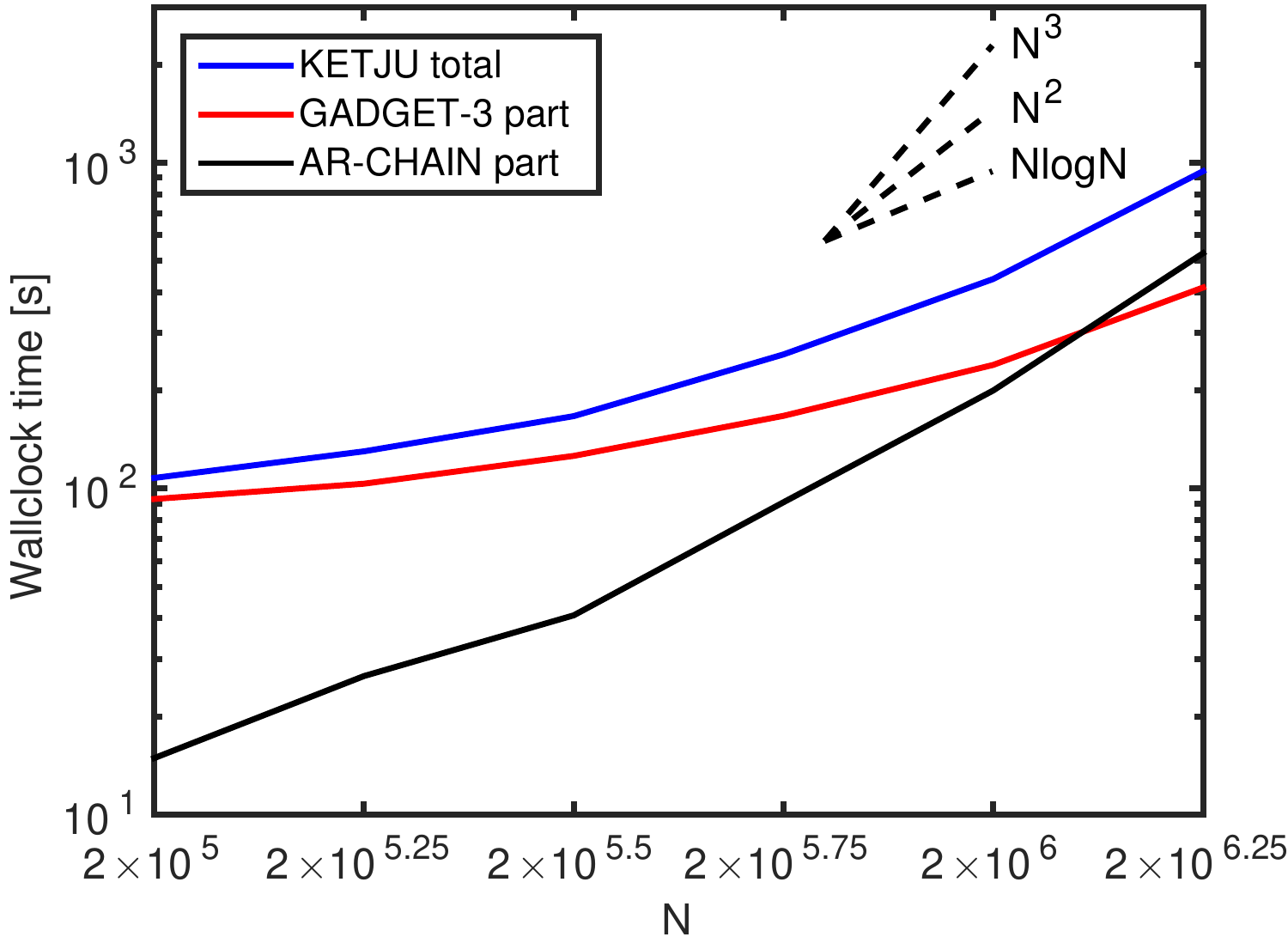}
\caption{The scaling test T1. At low particle numbers, the \gadget{} part 
dominates the time consumption of \ketju{} and the scaling is close to the 
characteristic $\bigO(N\log{N})$ scaling of \gadget. At $N \gg 10^6$ the 
\archain{} part dominates and the scaling is steeper. The maximum 
stellar particle number which can be run using the current version of 
\ketju{} within reasonable wallclock time is $N \sim 5 \times 10^6$.}
\label{fig: weak}
\end{figure}

We perform two scaling tests to study the performance of \ketju. The first 
scaling test T1 is run using a constant number of MPI tasks while modifying the 
number of particles in the initial conditions. In the second scaling test T2 
the test problem remains fixed but the number of MPI tasks is varied. For both 
of 
these test setups we use galaxy merger initial conditions containing two 
multi-component 
galaxy models with a stellar bulge, a dark matter halo and a central SMBH. 
For a detailed description 
of the initial setup, see \S \ref{multi_component} and Tables \ref{table: 
progenitors} and \ref{table: runs}. We set the chain radius to 18 pc 
$(\lambda=1.8$, $\gamma = 25)$ in all the scaling test runs. The DM softening 
length is set to $\epsilon_{\mathrm{DM}} = 100$ pc, and the stellar softening 
length is $\epsilon_\star = 6$ pc. The number of dark matter particles remains 
fixed at $N_{\mathrm{DM}} = 10^6$ in all the scaling test runs. For the scaling 
tests T1, we select the following accuracy parameters: the \gadget{} error 
tolerance $\eta$ $=0.002$ and the \archain{} GBS tolerance of 
$\eta_{\mathrm{GBS}} = 10^{-6}$. We use 96 MPI tasks in all the runs for test 
sample T1. The number of stellar particles is varied between $2\times 10^5 \leq 
N \leq 2\times 10^{6.25}$. The results of the scaling test T1 are presented in 
Fig. \ref{fig: weak}.

When the total number of particles is $N<5\times10^5$, the time consumption of 
\archain{} is negligible. As ordinary \gadget{} scales as $\bigO(N\log{N})$ 
and the scaling of \archain{} is steeper ($\sim \bigO(\tilde{N}^2)$), the 
wallclock time 
consumed by the chain computation will eventually exceed the time consumed by 
the 
\gadget{} part. This fact sets the limit on how large particle numbers \ketju{} 
can handle: it is not meaningful to run simulations in which the subsystem 
computations take most of the wallclock time. With the initial condition used 
here, the \archain{} part takes approximatively half of the computation time 
for 
$N \sim 2.5\times 10^6$ simulation particles.

\begin{figure}[h!]
\includegraphics[width=\linewidth]{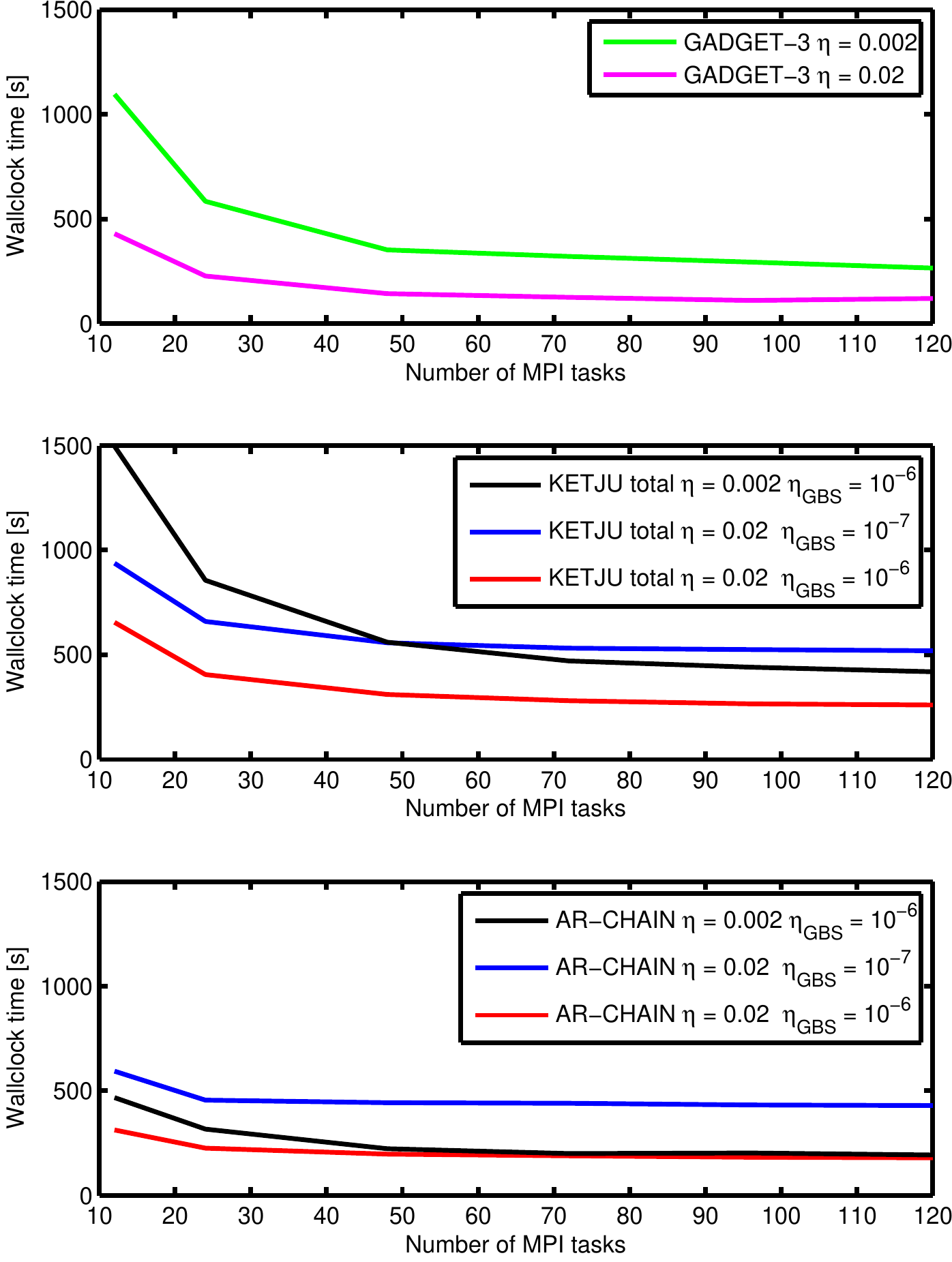}
\caption{The scaling test T2: the performance of \ketju{} and \gadget{} with  
different numbers of MPI tasks and different accuracy parameters. Top panel: 
the 
scaling for standard \gadget{}. In this particular test the code scales well up 
to $\sim 
50-75$ tasks after which the scaling is poor. Middle panel: the scaling of 
\ketju{} is very similar to standard \gadget, but \ketju{} consumes 
approximatively 50 \% more 
computational time. Bottom panel: the scaling of the \archain{} 
part is approximately flat, as expected. For more information about the 
node-based 
computation strategy, see the text. }
\label{fig: strong}
\end{figure}

In the second scaling test T2, we test the scaling of \ketju{} as a function 
of the number of MPI tasks. The number of stellar particles is fixed to 
$N=2\times10^6$. The results are shown in Fig. \ref{fig: strong}. We conclude 
that \ketju{} scales in a similar manner to standard \gadget: the scaling is 
good up to 
$\sim 50-75$ MPI tasks after which it is considerably worse. The value of the 
GBS accuracy parameter \etabs{} has a large effect on the \archain{} 
computational 
time. With \etabs{} $=10^{-6}$ the chain integration is performed $\sim 2.2$ 
times 
faster than when using \etabs $=10^{-7}$ with negligible differences in the 
results as can be seen in \S \ref{Test_code}. With the accuracy parameters in 
use  
in \S \ref{results} ($\eta = 0.002$, \etabs $=10^{-6}$) \ketju{} consumes 
roughly 50 \% more computational time than the standard \gadget.

From the scalability and timing tests we conclude that the \ketju{}
code is a fast energy-conserving regularized tree code suitable for simulations
with up to $N \sim 5 \times 10^6$ stellar particles in its current 
configuration. This 
is sufficient 
for our current purposes of simulating regularized isolated galaxies and galaxy 
mergers with 
SMBHs. We leave further 
code optimization, required for simulations with particle numbers in excess of 
$N > 5\times 10^6$, for future work. We also stress that our code improves on 
the maximum 
particle numbers used in the field of studying regularized SMBH dynamics in a 
galactic environment as current state-of-the-art NBODY simulations typically 
reach up to $5\times 10^5$ - $10^6$ simulation particles (see e.g. 
\citealt{wang2014} and \citealt{khan2011}), roughly a factor of $5-10$ below 
our highest resolution runs presented in \S \ref{results}.

\section{Results}
\label{results}

Numerical simulations of merging galaxies using direct summation codes with 
particle numbers of $N \leq 10^6$ have recently begun to establish a 
consensus that the final-parsec problem is in fact nonexistent in galaxy 
mergers 
\citep{khan2011,Preto2011}.
The main conclusion of these studies, that the SMBH hardening rate is in fact 
resolution-independent in galaxy mergers, was based on the fact that two-body 
relaxation is not driving the SMBH loss cone refilling. Instead, the 
non-spherical shape of the 
galaxy potential provides an additional torque on the stellar orbits, which 
fill 
the loss cone on a timescale much shorter than the two-body relaxation 
timescale. The hardening rate is also large enough to drive the binary to the 
gravitational wave dominated regime on a timescale that is short compared to 
the 
Hubble time.

In this section we use \ketju{} to study the resolution-dependence of the SMBH 
binary hardening rates and the timescales of the supermassive black hole 
mergers 
using two different types of initial conditions. The first 
type consists of a stellar bulge and a SMBH without a dark matter halo, a setup 
which has been extensively used in previous SMBH hardening rate studies (e.g. 
\citealt{khan2011, Preto2011}). In the second type of initial conditions 
we include a dark matter halo in addition to the stellar 
bulge and the SMBH components. The SMBH hardening rates in these more realistic 
multi-component models have not thus far been rigorously studied in the 
literature. 

\subsection{Multi-component equilibrium initial conditions}
\label{multi_component}

The initial conditions are generated here using the distribution 
function method (see e.g. \citealt{Merritt1985} and \citealt{Ciotti1992}). We 
model a typical massive elliptical galaxy as an isotropic, 
spherically symmetric multi-component Hernquist sphere consisting of three 
components: 
a stellar component, a dark matter halo and a central SMBH. For a single mass 
component $i$ the Hernquist density profile with mass $M_i$ and scale radius 
$a_i$ is defined 
as 
\begin{equation}
	\rho_i(r) = \frac{M_i}{2 \pi} \frac{a_i}{r\left( r + a_i\right)^3},
\end{equation}
which corresponds to the simple softened gravitational potential
\begin{equation}
	\Phi_i(r) = - \frac{G M_i}{r + a_i} 
\end{equation}
and the cumulative mass profile 
\begin{equation}\label{eq: massprofile}
	M_i(r) = M_i \frac{r^2}{\left(r + a_i \right)^2}. 
\end{equation}
The total multi-component potential $\Phi_{\mathrm{T}}$ is the sum of the 
stellar, 
dark matter and central SMBH potentials, and is parametrized in our 
implementation as
\begin{equation}
\begin{aligned}\label{eq: multipotential}
	\Phi_{\mathrm{T}} = \Phi_{\star} + \Phi_{\mathrm{DM}} +\Phi_{\bullet}\\
	 = - \frac{G M_\star}{r + a_{\star}} - \frac{G M_\mathrm{DM}}{r + 
a_\mathrm{DM}} - \frac{G M_\bullet}{r + \xi_{\bullet}}\\
	 = -G  M_\star \left[ \frac{1}{r+a_\star}  + \frac{\mu}{r+\beta 
a_\star} 
+ \frac{\eta}{r+ \xi_{\bullet} }      \right], 
\end{aligned}
\end{equation}
where the multi-component model parameters are defined as 
$\mu = M_{\mathrm{DM}} / M_\star$, $\eta = M_{\bullet} / M_{\star}$ and $\beta 
= 
a_{\mathrm{DM}} / a_\star$. This formulation extends the 
two-component parametrization of \cite{hilz2012} with the addition of the 
central SMBH 
parameters. For numerical reasons, the SMBH potential is softened with a small 
gravitational softening length of $\xi_{\bullet}$ in order to ensure that the 
total potential remains finite at $r=0$. We set $\xi_\bullet = 1 \times 
10^{-5}$ kpc. Note that $\xi_{\bullet}$ should not be confused with the 
gravitational softening length of the SMBH in the tree code.

The velocity profiles for the stellar and DM components are obtained using 
their respective phase-space distribution functions $f_i$. This approach has 
the advantage that it results in more stable initial conditions than using 
Jeans equations \citep{binney2008, kazan2004}. In general, the distribution 
function $f_i$ of a density component $\rho_i$ in the total gravitational 
potential $\Phi_{\mathrm{T}}$ is computed using 
Eddington's formula \citep{binney2008}:

\begin{equation}
	f_i \left(\mathcal{E} \right) = \frac{1}{\sqrt{8} \pi^2} 
\int_{\Phi_{\mathrm{T}} = 0}^{\Phi_{\mathrm{T}} = \mathcal{E}} \frac{d^2 
\rho_i}{d \Phi_{\mathrm{T}}^2} \frac{d 
\Phi_{\mathrm{T}}}{\sqrt{\mathcal{E}-\Phi_{\mathrm{T}}}},
\label{eq: eddington}
\end{equation}
in which $\mathcal{E} = -\frac{1}{2} v^2 - \Phi_{\mathrm{T}} + \Phi_0$ is the 
(positive) energy relative to the chosen zero point of the potential $\Phi_0$. 
In general, the zero point is chosen so that $f_i > 0$ for $\mathcal{E} > 0$ 
and $f_i = 0$ for $\mathcal{E} \le 0$. For an isolated system extending to 
infinity, such as the 
Hernquist 
sphere, we set $\Phi_0 = 0$.
Unfortunately, the term $\rho_i(\Phi_{\mathrm{T}})$ does not have an analytical 
expression in the general case. Therefore we rewrite the 
derivative term of Eq. \eqref{eq: eddington} using the chain rule, following 
\cite{hilz2012}:

\begin{equation}\label{eq: edd-integrand}
\frac{d^2 
\rho_i}{d \Phi_{\mathrm{T}}^2} = 
\left(\frac{d \Phi_{\mathrm{T}}}{dr}\right)^{-2} \left[ \frac{d^2\rho_i}{dr^2} 
- 
\left(\frac{d \Phi_{\mathrm{T}}}{dr}\right)^{-1}  \frac{d^2 \Phi_{\mathrm{T}}   
}{dr^2} \frac{d \rho_i}{dr}      \right].
\end{equation}
The second term of the integral in Eq. \eqref{eq: eddington} is simply
\begin{equation}\label{eq: edd-integrand2}
\frac{d \Phi_{\mathrm{T}}}{\sqrt{\mathcal{E}-\Phi_{\mathrm{T}}}}  = \frac{d 
\Phi_{\mathrm{T}}   }{dr} 
\frac{dr}{\sqrt{\mathcal{E}-\Phi_{\mathrm{T}}}}.
\end{equation}
The resulting expressions contain only first and second derivatives of 
the density $\rho_i$ and the total potential $\Phi_{\mathrm{T}}$ with respect 
to $r$, which are easily obtained by taking the derivatives of their analytical 
formulas. Using $r$ as the 
integration variable naturally changes the limits of the integration. 
$\Phi_{\mathrm{T}}(r) = \mathcal{E}$ has 
to be inverted numerically for $r$, whereas the lower integration limit 
$\Phi_{\mathrm{T}}(r) = 0$ corresponds 
to $r = \infty$. 

We compute a random realization of a multi-component Hernquist sphere 
using the following procedure. First, we draw the random particle positions for 
the stellar and dark matter components using the inverse cumulative mass 
profile 
from Eq. \eqref{eq: massprofile}. Next, we compute the values of the 
distribution 
functions $f_\star(\mathcal{E})$ and $f_{\mathrm{DM}}(\mathcal{E})$ into a 
lookup table 
using Eqs. \eqref{eq: eddington} and \eqref{eq: edd-integrand}. After this we 
sample the random particle velocities in a computationally efficient way by
interpolating the tabulated values of the distribution functions. Finally, we 
place a SMBH at rest at the center of the multi-component sphere. 

We also note here that our initial conditions assume no 
gravitational softening. Taking the non-zero gravitational softening length 
into account would result in even more stable initial conditions. Initial
conditions that compensate for the gravitational softening
have been introduced by e.g. \cite{Muzzio2005} and \cite{Barnes2012}. However, 
in \ketju, 
the innermost region of the galaxy potential around the central SMBH within the 
chain radius 
$r_{\mathbf{chain}}$ is not softened while the rest of the potential is. 
Consequently, implementing 
the softening correction in the IC generation may not be completely 
straightforward and is left as a topic for future code development.

\subsection{Galaxy models}
\label{galaxy_models}

We use two principal types of galaxy models in this study:
two-component models that in addition to the SMBH only include a stellar bulge 
(B sample) and 
three-component models that include a DM
halo in addition to the stellar bulge and the central SMBH (H sample). For the 
stellar bulge component, we set $M_\star = 10^{11} 
M_\odot$ and $a_\star = 1.5$ kpc, motivated by the observations of 
\cite{vanderwel2014} of the mass-size relation of $z \sim 1$ massive early-type 
galaxies. For the multi-component models including a  DM halo, we set, 
following 
\citet{hilz2012}, $\beta = 11$ and $\mu = 100$ motivated by the halo abundance 
matching results of \cite{moster2013}  yielding 
$M_{\mathrm{DM}} = 10^{13} M_\odot$ 
and $a_{\mathrm{DM}} = 16.5$ kpc for the dark matter component. The mass of the 
central SMBH is set to $M_\bullet = 10^8 M_\odot$ resulting in $\eta = 
M_\bullet 
/ 
M_\star = 0.001$, see Table \ref{table: progenitors}. 
The motivation for setting up these two simulations samples was to study the 
hardening rate of the black hole binary in a purely baroynic setting (B sample) 
and in a 
setting with a high dark matter fraction (H sample). In this way our two 
simulation samples 
bracket the environments found in the centers of typical elliptical galaxies.

\begin{figure}[h]
\includegraphics[width=\linewidth]{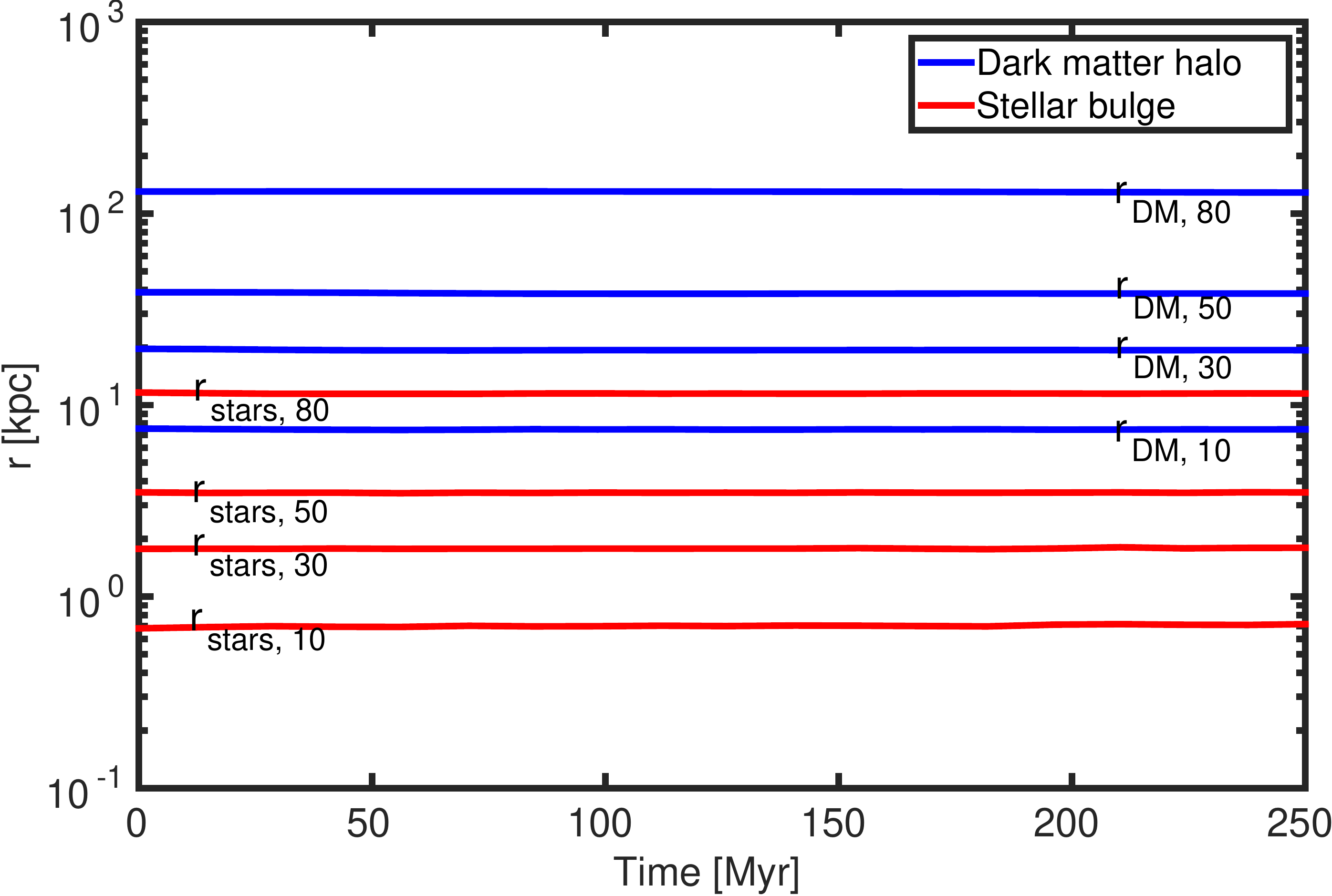}
\caption{The stability of the mass radii of an isolated three-component galaxy. 
The 
smallest stellar mass radius, $r_{10}$, enclosing the innermost 10\% of the 
stellar mass 
increases by $\sim 3$\% during the run. Other stellar and DM radii remain even 
more stable.}
\label{fig: hilz_stability}
\end{figure}

The stability of a three-component model with particle numbers 
($N_{\mathrm{DM}} = 10^6$, $N_\star = 10^7$, a single central SMBH) is studied 
according to the stability 
test of \cite{hilz2012} using the standard \gadget \ code with a gravitational 
softening length of $\epsilon=20$ pc. The results of the stability 
test are presented in Fig. \ref{fig: hilz_stability}. The radii containing 
10\%, 30\%, 50\% and 80\% of the total dark matter mass remain within 1\% of 
their original values during the entire simulation timespan of $t=250$ Myr 
which corresponds roughly to $\sim 80$ dynamical time scales at the radius 
enclosing $10$ \% of the total stellar mass. The 10\% stellar mass radii 
increases by $\sim 3$ \% during the simulation, whereas the other stellar mass 
radii show 
even less variation, thus validating the stability of our three-component 
initial conditions.\\

The two-body relaxation time scale $t_{\mathrm{relax}}$ is defined as
\begin{equation} 
t_{\mathrm{relax}} \approx \frac{0.1 N}{\ln{N}} \times 
t_{\mathrm{cross}},
\end{equation}
where $N$ is the number of particles and $ t_{\mathrm{cross}}$ is the crossing 
time \citep{binney2008}. As the number of stars $(N\sim 10^{11})$ in a massive 
elliptical galaxy still exceeds the particle number in modern galactic-scale 
numerical simulations by several orders of magnitude, the simulated galaxies 
are 
subject to spurious two-body relaxation effects in their very  dense central 
regions (e.g. \citealt{Diemand2004}), if no or very small gravitational 
softening is used. This usually results in a core-like structure in the central 
region, even without a SMBH, as stars are scattered to lower binding energies 
\citep{hilz2012}.\\

\subsection{Galaxy mergers with SMBHs}

\begin{table}
\caption{Properties of merger progenitor galaxies}
    \begin{tabular}{| c | c | c | c | c | c |}
    \hline
    Simulation sample & $M_\star [10^{10} M_\odot]$  & $\mu$  & $\eta$ & 
$a_\star$ [kpc] & $\beta$\\ 
\hline
    B & $10^{11}$ & - & $0.001$ & 1.5 & -\\
    H & $10^{11}$ & $100$ & $0.001$ & 1.5 & 11\\
    \hline
    \end{tabular}
    \label{table: progenitors}
\end{table}

We set up a sample of 57 major galaxy mergers using the progenitors 
described 
in the previous section and in Table \ref{table: progenitors}. The simulation
sample is run using \ketju \ in order to study the dependence of the SMBH 
binary hardening and the eccentricity evolution on 
the adopted stellar mass resolution. First, we focus on the binary hardening 
phase dominated by three-body interactions with the surrounding stars, where 
the 
semi-major axis of the binary $a$ lies within $0.1$ pc $ \lesssim a \lesssim 5 
\ 
\rm pc$ and the Post-Newtonian corrections can be safely neglected. 
In \S \ref{pn-section} we run a sample of high-resolution simulations including 
Post-Newtonian corrections 
until the SMBHs coalesce. A total number of 26 major mergers are run with 
initial 
conditions 
resembling the ones used in the earlier studies of \cite{khan2011} and 
\cite{Preto2011}: colliding massive stellar bulges without a dark matter halo 
(sample B in Table \ref{table: runs}). The rest of the simulations have a 
multi-component initial setup: the stellar bulges reside in massive 
dark matter halos (sample H in Table \ref{table: runs}). The particle numbers 
in different simulations are also presented in Table \ref{table: runs} and 
range 
from $2\times 10^{5}$ 
stellar particles (simulations B1 \& H1) to the maximum particle number 
used in this 
study: $2\times10^6$ DM particles 
and $2\times10^{6.25} \approx 3.6\times10^6$ stellar particles in the merger 
remnant (simulations B6 \& H6). The different simulations within each 
set for a given 
resolution only differ in the random seed used in setting up the initial 
conditions. All simulation samples except for the sample H5 PN are fully 
Newtonian.\\

\begin{table}
\caption{Particle numbers in the merger remnants}
    \begin{tabular}{| c | l | l | c |}
    \hline
    Sample label & $N_{\star}$ & $N_{\mathrm{DM}}$ & Number of runs with\\ 
& & & different random seeds\\  \hline
    B1 & $2\times10^{5}$ & - &5 \\
    B2 & $2\times10^{5.25}$ & - &5 \\
    B3 & $2\times10^{5.5}$ & - &5 \\
    B4 & $2\times10^{5.75}$ & - &5 \\
    B5 & $2\times10^{6}$ & - &5 \\
    B6 & $2\times10^{6.25}$ & - &1 \\
    H1 & $2\times10^{5}$ & $2\times10^{6}$ & 5 \\
    H2 & $2\times10^{5.25}$ & $2\times10^{6}$ & 5 \\
    H3 & $2\times10^{5.5}$ & $2\times10^{6}$ & 5 \\
    H4 & $2\times10^{5.75}$ & $2\times10^{6}$ & 5\\
    H5 & $2\times10^{6}$ & $2\times10^{6}$ & 5\\
    H6 & $2\times10^{6.25}$ & $2\times10^{6}$ & 1 \\
    H5 PN & $2\times10^{6}$ & $2\times10^{6}$ & 5\\
    \hline
    \end{tabular}
\label{table: runs}
\end{table}   

The bulge-only galaxies in runs B1-B6 are set on the merger orbit in the 
following way. The initial separation is chosen to be $d=20\times a_\star = 30$ 
kpc. The encounter orbits are nearly parabolic as motivated by cosmological 
simulations \citep{Khochfar2006} with the initial velocities chosen as such 
that 
the separation of the galactic nuclei is approximately the scale radius 
$a_\star$ during the first 
pericenter passage. 
The initial velocities of the galaxies in the H sample (H1-H6) are defined in a 
slightly different manner. 
For these mergers, we only consider the mass inside $d=30$ kpc when computing 
the 
initial velocities of 
the galaxies on the parabolic orbit. This results in a slightly faster 
coalescence of the galactic nuclei in the H 
sample runs compared to B sample runs.
The chain radius is set to $18 \ \rm pc$ in all the simulation runs by 
choosing chain parameters $\lambda = 1.8$ and 
$\gamma = 25$. For all runs the gravitational softening lengths are set to 
$\epsilon_\star = 
6$ pc and $\epsilon_{\mathrm{DM}} = 100$ pc for the stellar and dark matter 
components, respectively.
In order to ensure the accuracy of the code with small gravitational softening 
lengths, we also 
set the \gadget{} integrator error tolerance parameter to $\eta = 0.002$, which 
is smaller 
than the canonical \gadget{} parameter value by a factor of $\sim 10$. We also 
tested 
the chosen set of code parameters for potential pathological simulation 
behavior. Several 
runs with slightly different softening lengths, chain radii and error tolerance 
parameters were performed with similar results compared to the runs with our 
chosen standard code parameters.

\begin{figure}
\includegraphics[width=\linewidth]{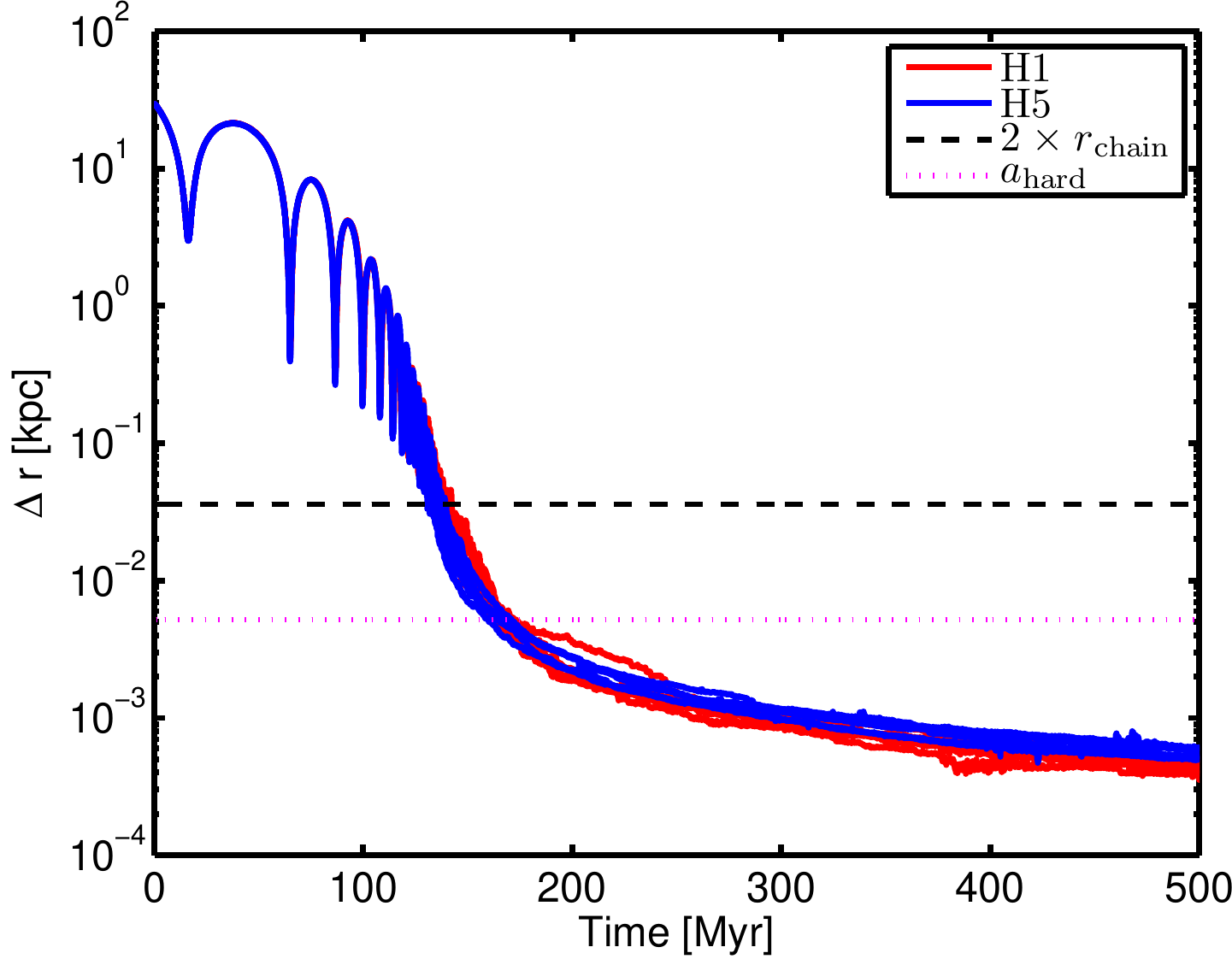}
\caption{The smoothed SMBH separation in the simulation runs H1 and H5. 
The SMBHs form a bound binary at $t\sim 134$ Myr. At this stage the respective
stellar cusps of the SMBHs are disrupted and the subsequent hardening of the 
binary orbit proceeds through three-body interactions. The dashed line marks 
the radial separation after which the SMBHs belong to the same 
chain subsystem. The dotted line depicts the the semi-major axis of a hard SMBH 
binary, as decribed in the main text.}
\label{fig: khan_separation}
\end{figure}

\begin{figure*}
\includegraphics[width=\textwidth]{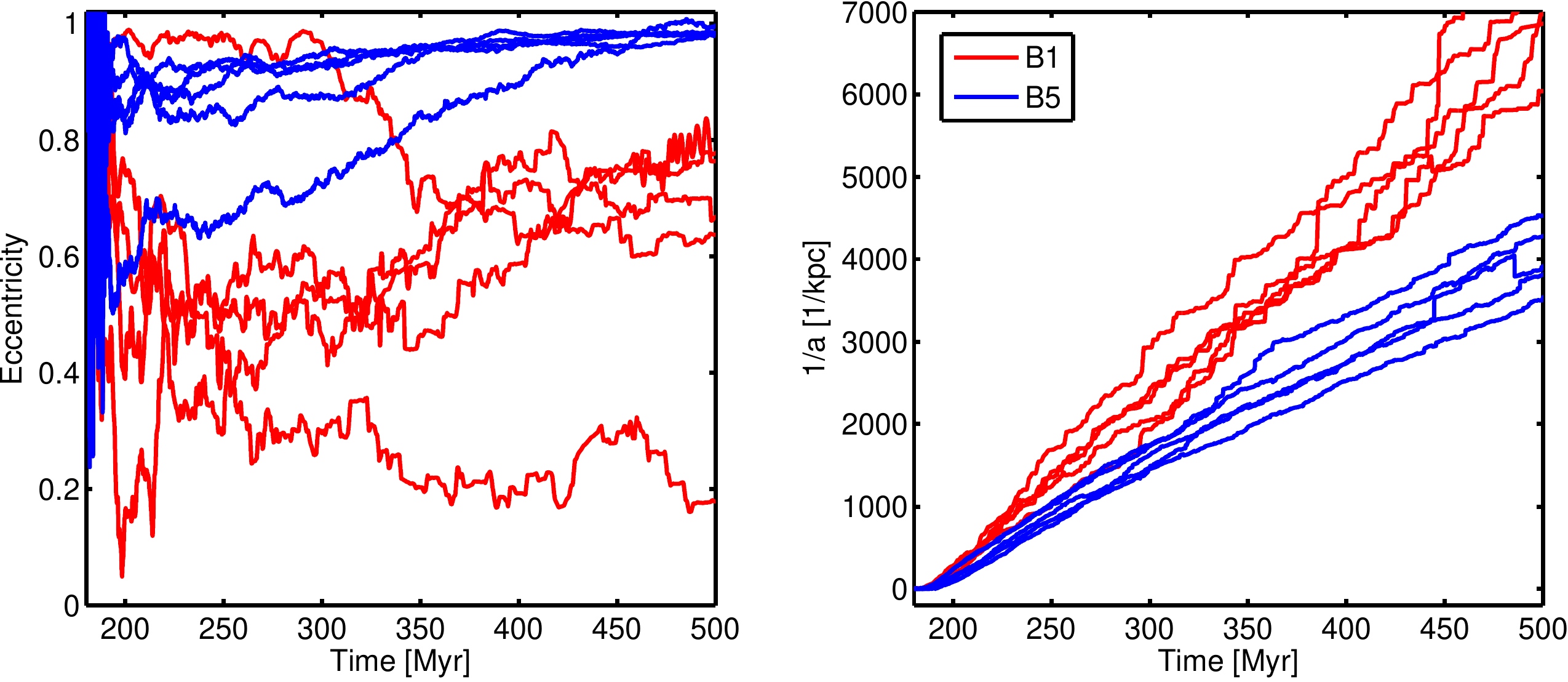}
\caption{Binary evolution in the bulge-only simulation sample B. The 
lines with the same color have the same mass resolution but different random 
seeds in their initial conditions.} Left panel: 
the eccentricity evolution after the binary formation. The binary eccentricity 
is clearly higher and more converged in the high-resolution runs. Right panel: 
the inverse semi-major axis. The hardening rate $d(1/a)/dt$ decreases when 
going 
from low to high resolutions, as expected.
\label{fig: khan_hardening_B}
\end{figure*}

\subsection{SMBH binary evolution in merging galaxies}\label{binaryevolution}

We run all the merger simulations for $t=500$ Myr. The SMBHs lie within 
the central cusps of their host galaxies for several close passages of the 
galactic nuclei during the 
merger until the cusps merge and are quickly disrupted by the formation of the 
SMBH binary. This occurs at $t \sim 134$ Myr in sample H and at $t \sim 184$ 
Myr in simulation sample B due to the different encounter orbits of the 
galaxies with and without the DM halo. After the disruption of the stellar 
cusps, the 
dynamical friction becomes inefficient and the binary subsequently hardens via 
three-body interactions with the surrounding stars. The binary becomes hard 
when its semi-major axis satisfies the criterion $a<a_\mathrm{hard}$. We adopt 
the definition used by \cite{Merritt2005} and \cite{Merritt2007}: for an 
equal-mass SMBH binary $a_\mathrm{hard} = r_\mathrm{infl}/16$, in which the 
influence radius $r_\mathrm{infl}$ is the radius enclosing stellar mass 
$M_\star(r_\mathrm{infl}) = 2 M_\bullet$. With this definition, 
$a_\mathrm{hard} 
\sim 5$ pc all the runs.  The other commonly used definition for a hard binary 
is $a_\mathrm{hard} = G \mu/4\sigma_\star^2$, where $\mu$ is the reduced 
mass of the SMBH binary and $\sigma_\star$ is the nuclear stellar velocity 
dispersion \citep{Merritt2006}. With this definition $a_\mathrm{hard}$ gets 
slightly lower values. The slingshot-hardening phase continues until the 
simulation end time at $t=500$ Myr, after which the semi-major axis of 
the SMBH binary is $0.1$ pc $\lesssim a \lesssim 0.4$ pc $< a_{\mathrm{hard}}$ 
pc in all the simulation runs. The separation of the two SMBHs as a 
function of time during the galaxy merger for simulations H1 and H5 is 
presented 
in Fig. \ref{fig: khan_separation}.

\begin{figure*}
\includegraphics[width=\textwidth]{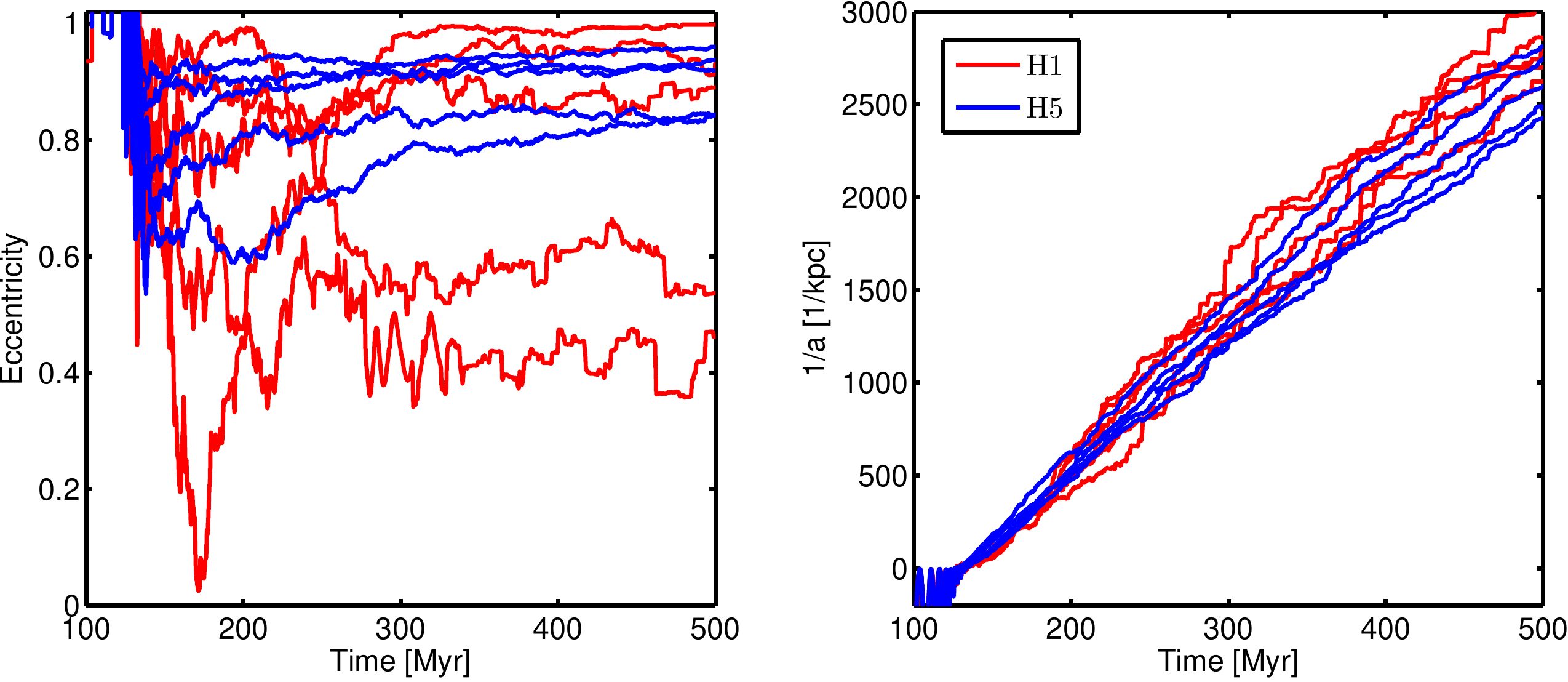}
\caption{
The evolution of the binary in the halo-included simulation sample H. The lines 
with the same color have the same mass resolution but 
different random seeds in their initial conditions. The binary eccentricities 
are quantitatively similar in simulation samples B and H. However, in contrast
to simulation sample B the evolution of the hardening rate $d(1/a)/dt$ of the
 inverse semi-major axis for sample H shows no apparent resolution-dependence.}
\label{fig: khan_hardening_H}
\end{figure*}

We show the evolution of the orbital eccentricity and the inverse 
semi-major axis of the SMBH binaries of sample B1 and B5 in Fig. \ref{fig: 
khan_hardening_B} and H1 and H5 in Fig. \ref{fig: khan_hardening_H}. 
The binary eccentricities in the high-resolution simulations B5 and H5 are in 
general high, $0.6 < e < 1.0$ with most of the binaries having eccentricities 
in 
excess of $e > 0.8$. At 
lower resolutions, the stellar field surrounding the binary is resolved less 
accurately resulting in lower binary eccentricities and a significantly larger 
scatter between different runs with different initial random seeds. This is 
expected since the eccentricity evolution of the binary, especially at the 
moment of the binary formation, depends sensitively on the velocity field of 
the 
surrounding stars \citep{mikkola1992}. The eccentricity of the SMBH 
binary also increases due to the exchange of angular momentum with the 
surrounding stars. Typically, the SMBH binary eccentricity increases during the 
slingshot-hardening phase if it is high enough to begin with. However, the 
eccentricity growth rate 
decreases for binaries with initially low eccentricities $(e < 0.5)$ and is 
$\sim 0$ for circular 
binaries \citep{Sesana2006}. 
This  situation may be different in stellar systems that are strongly corotating 
with the SMBH 
binary \citep{Sesana2011} as the binary circularizes instead of becoming more 
eccentric.

\begin{figure}[h!]
\includegraphics[width=\linewidth]{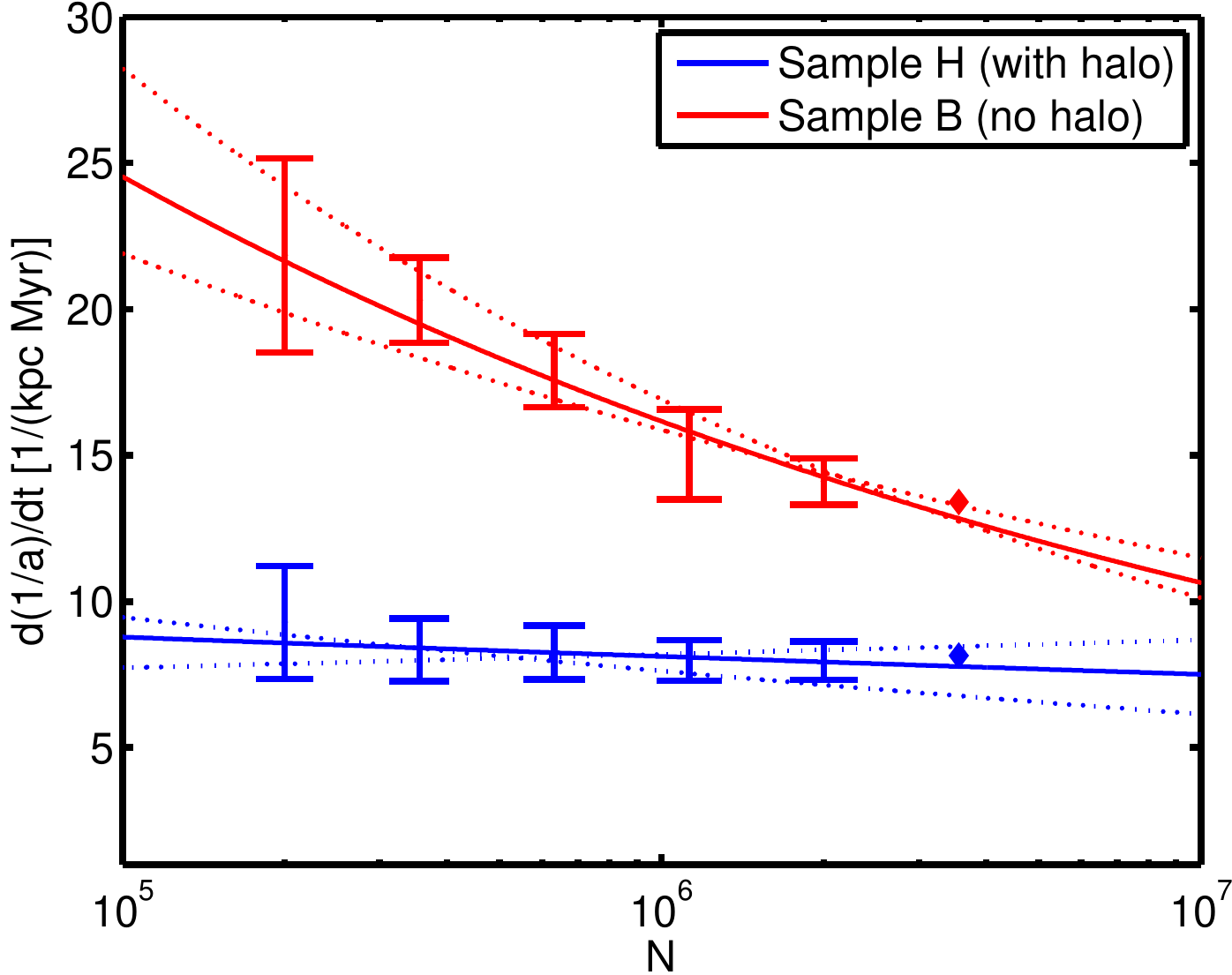}
\caption{The SMBH binary hardening rates as function of the simulation 
stellar 
particle number, represented with error bars of 1 standard deviation. The 
bulge-only simulations show a dependence on the particle number, $\propto 
N^{-0.18 \pm 0.04}$. If the DM halo is included, the power-law $\propto 
N^{-0.03 
\pm 0.06}$ is 
consistent with no resolution-dependence in the H sample, as described in the 
text. The three fits for each sample are plotted with the mean power-law 
exponent and the mean exponent $\pm$ error of one standard deviation.\\}
\label{fig: khan_scaling}
\end{figure}

The evolution of the inverse semi-major axis is very close to linear 
($1/a \propto t$) during the hardening phase in all the runs of simulation 
samples B 
and H until $t = 500$ Myr. The binary hardening rates are presented in Fig. 
\ref{fig: khan_scaling} as a function of the stellar particle number of the 
merger remnant. The mean hardening rates with errors of one standard deviation 
from the selected samples are as follows: B1: $21.8 \pm 3.3$ 
kpc$^{-1}$Myr$^{-1}$, B5: $14.1 \pm 0.8$ kpc$^{-1}$Myr$^{-1}$, H1: $9.3 \pm 
1.9$ 
kpc$^{-1}$ Myr$^{-1}$ and H5: $8.0 \pm 0.7$ kpc$^{-1}$Myr$^{-1}$. We further 
quantify the resolution dependence of the hardening rates by fitting a 
power-law 
$d(1/a)/dt \propto N^{-\alpha}$ to the results and studying the distribution of 
the power-law exponent $\alpha$. This is done by using a simple bootstrap 
method. Considering first sample B, we pick a random run from each subsample 
B1-B6 each and fit the power-law. We repeat this procedure $10^4$ times and 
obtain the mean $\alpha$ and its standard deviation. The process is then 
repeated for sample H. In sample B, the hardening rate clearly depends on 
the stellar mass resolution:
\begin{equation}
  \derfrac{}{t}\left(\frac{1}{a}\right)_{\mathrm{B}} \propto N^{-0.18 \pm 0.04},
\end{equation}
whereas, sample H is consistent with no resolution dependence of the 
hardening rate:
\begin{equation}
  \derfrac{}{t}\left(\frac{1}{a}\right)_\mathrm{H} \propto N^{-0.03 \pm 0.06}.
\end{equation}

\begin{figure*}
\includegraphics[width=\textwidth]{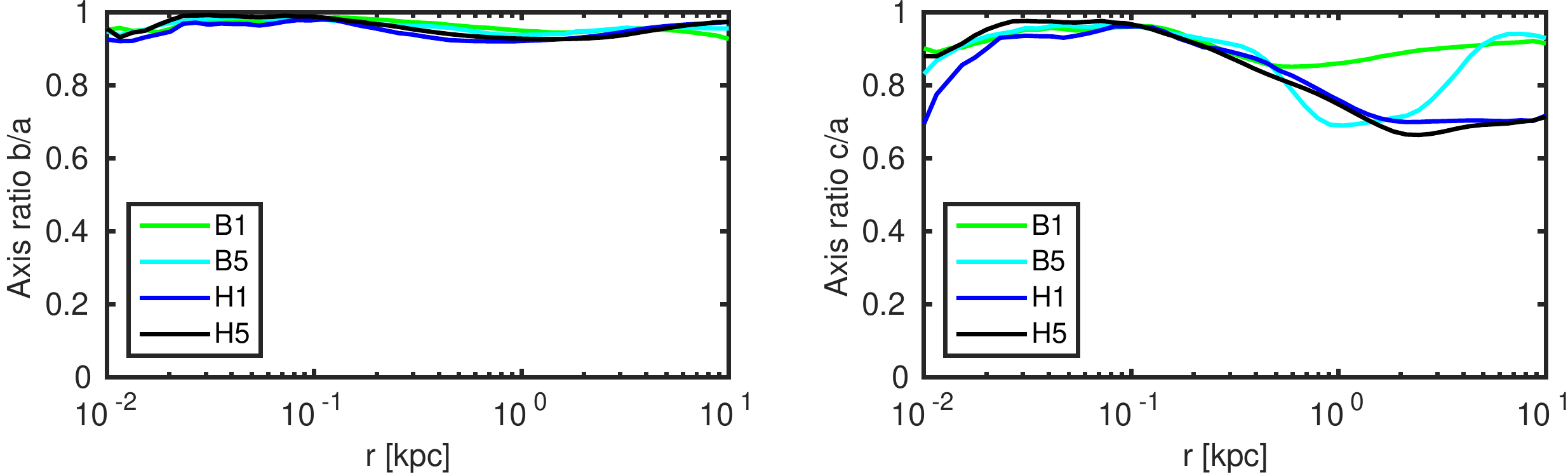}
\caption{The host galaxy axis ratios $b/a$ and $c/a$ in merger 
samples B5 and H5 as a function of radius. Left panel: the ratio $b/a$ is high, 
$\sim 0.95$, in all simulation samples B1, B5, H1 and H5. Right panel: the axis 
ratio $c/a$. The axis ratios are similar for samples H1 and H5, $c/a \sim 0.7$ 
at $r>1$ kpc, with $c/a$ close to unity inside this radius. However, the 
$c/a$ values for B1 and B5 clearly differ from each other and the H sample 
simulations, see text for further details.}
\label{fig: remnant_shape}
\end{figure*}

The resolution-dependence of the hardening rate originates from the 
resolution-dependence of the process which fills the loss cone of the SMBH 
binary. If the shape of the merger remnant is sufficiently asymmetric, the 
stellar orbits are torqued into the loss cone of the SMBH binary at a 
rate higher than the loss cone filling produced by the 
resolution-dependent two-body relaxation. Consequently, the 
resolution-dependence of the SMBH binary hardening rate decreases 
\citep{Merritt2004, berczik2006, khan2011}. We next study the shapes of the 
merger remnants in samples B and H. In calculating the shapes of the merger
remnants we closely follow the S1 method of \cite{Zemp2011}. The axis ratios 
$b/a$ and $c/a$ 
are computed in thin ellipsoidal shells from the eigenvalues of the shape 
tensor 
of the stellar 
matter distribution. We note that the axis ratios of the merger remnants remain 
roughly constant after the nuclei of the progenitor galaxies have merged. The 
axis ratios for the simulation samples B1, B5, H1 and H5 are presented in Fig. 
\ref{fig: remnant_shape} as a function of the distance from the center of the 
galaxy.

All the merger remnants are roughly axisymmetric: $b/a \sim 0.9$ for 
all the simulation samples between $10^{-2}$ kpc $ \leq r \leq 10$ kpc. The 
$c/a$ 
ratio is 
roughly $0.8$ near the SMBH binary and $\sim 0.95$ at $r=0.1$ kpc for all the 
simulation samples. However, at larger radii the differences between the 
samples 
become 
evident. For samples H1 and H5 the $c/a$ ratio decreases outwards and is $\sim 
0.7$ at $r=10$ kpc. For the sample B1 $0.8 < c/a < 0.9$ in the outer parts of 
the galaxy, whereas simulation sample B5 contains a flatter region with $c/a 
\sim 0.7$ 
between $0.5$ kpc $ \lesssim r \lesssim 5$ kpc. We attribute this phenomenon to 
the fact that relaxation effects are stronger in low-resolution simulations 
(e.g. \citealt{Power2003}), and the flatter feature has relaxed away in the B1 
lower resolution 
runs. The 
stellar orbits are defined by the total potential $\Phi_{\mathrm{T}}$. If the 
massive DM halo is present, the potential of the galaxy $\Phi_{\mathrm{T}}$ is 
dominated by the 
collisionless halo component $\Phi_{\mathrm{DM}}$ and the relaxation-induced 
evolution of the stellar component $\Phi_\star$ has only a small effect on the 
total potential $\Phi_{\mathrm{T}}$. Thus, it is natural that the axis ratios 
of samples H1 and H5 are more similar than for the bulge only B1 and 
B5 samples.

\subsection{Quantifying the differences in the hardening rates}

After studying the ellipsoidal axis ratios of the merger remnants, we 
further quantify the differences of the stellar populations in simulation 
samples B1, B5 and H5. Both the SMBH binary hardening hardening rates and 
merger remnant axis ratios in sample H1 are very close to the corresponding 
quantities of sample H5, thus we left sample H1 out of the analysis. In 
addition to the loss-cone filling rate, the hardening rate of the SMBH binary 
also depends on the distribution of the pericenter distances and the velocties 
of the incoming stars. Three-body scattering experiments have shown that, on 
average, stars with smaller pericenter distance and smaller initial velocity 
gain more energy from the binary in the star-binary interaction (see e.g.
\citealt{Valtonen2006} for a review on this topic.)

\begin{figure}
\includegraphics[width=\linewidth]{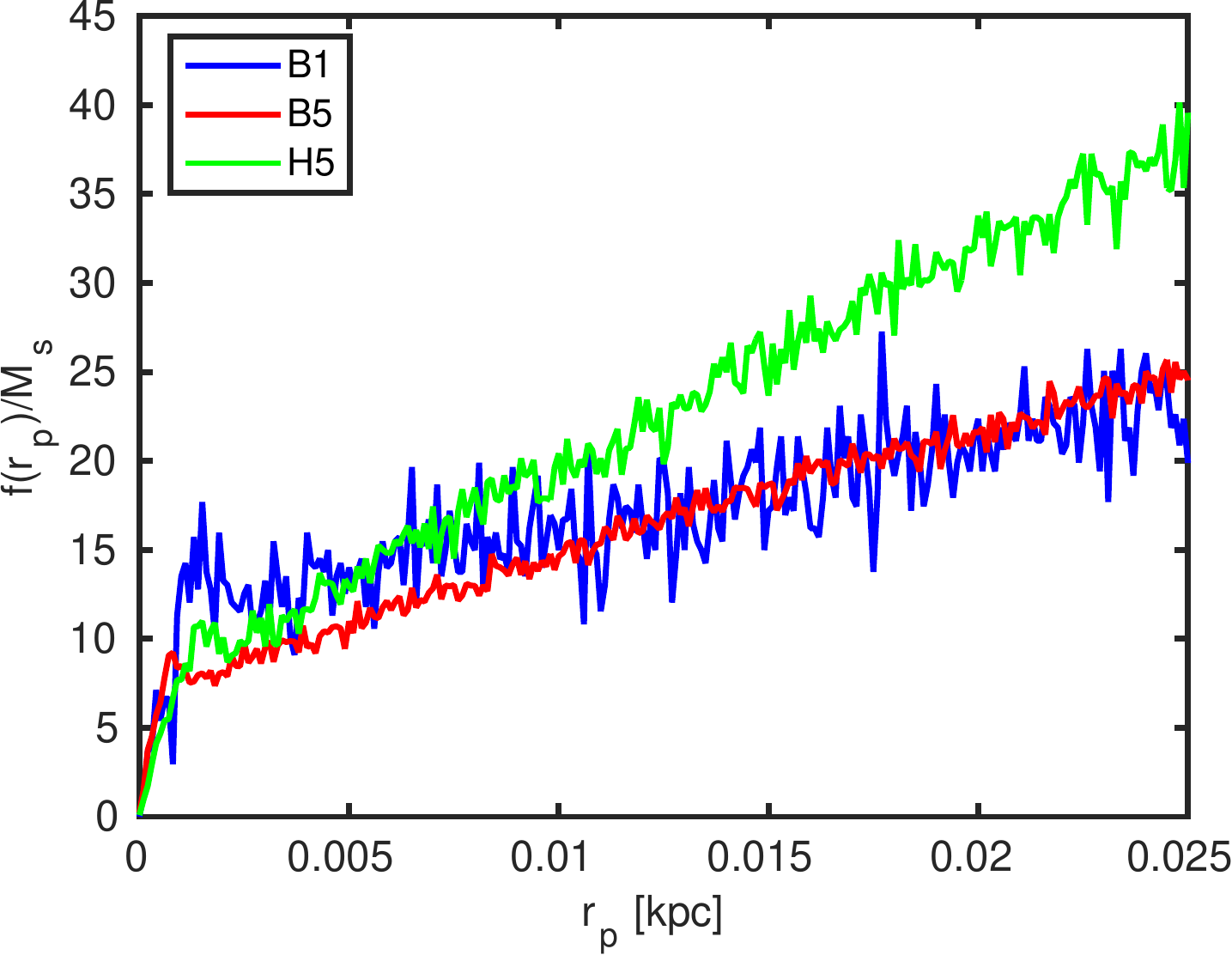}
\caption{The distribution of the stellar pericenter distances with respect to 
the center-of-mass of the SMBH binary during $200$ Myr $<t<400$ Myr. During 
this 
period, the 
semi-major axis of the binary is between $5$ pc and $0.5$ pc in all the 
simulation runs. The distribution function $f(r_{\mathrm{p}})$ is 
normalized to the total stellar mass $M_{\mathrm{s}}$ entering a sphere within 
a 
radius of 
$r = 30$ pc centered on the center-of-mass of the SMBH binary. In sample B1, a 
larger fraction of 
stars have pericenter distances of $r_{\mathrm{p}}<10$ pc compared to sample 
B5.}
\label{fig: rperi}
\end{figure}

We present the distribution of pericenter distances $r_\mathrm{p}$ of 
stellar particles in Fig. 
\ref{fig: rperi} during $200$ Myr $<t< 400$ Myr. During this period, the 
semi-major 
axis 
$a$ of the binary is between $5$ pc and $0.5$ pc in all the simulation runs. 
The pericenter distance distribution $f(r_\mathrm{p})$ is normalized by the 
total 
stellar mass $M_\mathrm{s}$ entering a sphere with a radius of $r=30$ pc 
centered on
the binary. The radius of $30$ pc is chosen based on the numerical criterion 
that stars with pericentric distances of $r_\mathrm{p} \lesssim 6\times a$ 
interact 
strongly with the binary \citep{mikkola1992}. Stars passing the binary with a 
larger pericenter 
distance just perturb the binary's center-of-mass and have a negligible effect 
on the 
orbital elements of the binary itself (e.g. \citealt{quinlan1996}). In sample 
H5, more 
stellar mass enters the vicinity of the SMBH binary compared to the samples B1 
and B5. This is due to the fact that the velocity dispersion for the H5 sample 
is higher than in the B simulation samples because of the presence of the 
massive DM halo. 
The additional incoming stars in H5 are distributed towards larger 
values of $r_\mathrm{p}$ compared to B1 and B5. However, a larger number of 
incoming stars does not 
make the hardening rates of SMBH binaries in the sample H5 greater, as also 
the velocities of the stars are higher.  
Comparing the bulge-only samples B1 
and B5 we see that there is more stellar mass at the pericenter distances 
of $r_\mathrm{p} < 10$ pc in the low-resolution run. Again, we attribute this 
phenomenon to spurious relaxation effects due to the low stellar mass 
resolution in the B1 sample. More stellar mass at low pericenter distances in 
sample B1 
contributes to the higher SMBH binary hardening rate compared to higher 
resolution sample B5.

\begin{figure*}
\includegraphics[width=\textwidth]{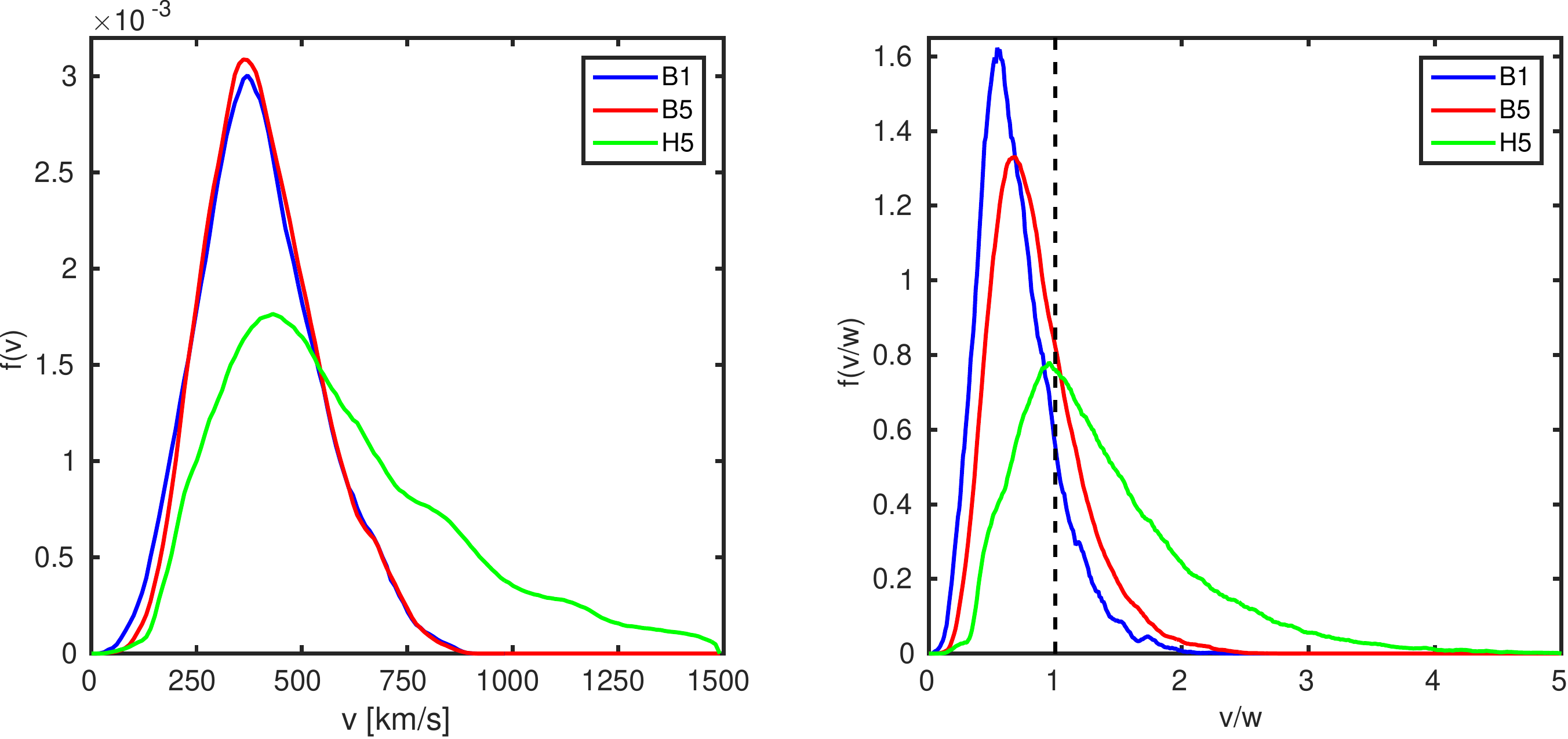}
\caption{Left panel: the velocity distribution of stellar particles crossing a 
shell at a distance
$r=30$ pc from the SMBH binary during $200$ Myr $<t<400$ Myr. Because 
of the deep potential well of the DM halo in H5, there are fewer particles with 
low velocities and a large number of particles with high velocities $v>750$ 
km/s 
compared to the bulge-only samples. The difference between the B1 and B5 
samples 
is the low-velocity tail in the distribution of B1. Right panel: the 
distribution of stellar particle velocities with respect to the time-dependent 
watershed velocity $w$, see the text for a definition. Now the difference of 
samples B1 
and B5 becomes clear: there are more particles with velocties below the 
watershed velocity $w$ in B1 
than in B5. These slowly moving stellar particles harden the SMBH binary 
efficiently.}
\label{fig: velocities}
\end{figure*}

Next, we study the distribution of the stellar particles which pass 
closer than $r=30$ pc from the center-of-mass of the binary in merger samples 
B1, B5 and 
H5. Here the concept of the so-called watershed velocity $w$ is extremely 
useful: 
\begin{equation}
 w \approx 0.85 \times \sqrt{\frac{M_{\bullet,1}}{M_{\bullet,1}+M_{\bullet,2}}} 
V_{\mathrm{bin}} = 0.85 \times \sqrt{ \frac{GM_{\bullet,2}}{a}},
\end{equation}
in which $V_\mathrm{bin}$ is the orbital velocity of the SMBH binary components 
\citep{quinlan1996}. For an equal-mass SMBH binary, the watershed velocity is 
$w \approx 0.6 \times V_\mathrm{bin}$.  On average, incoming stars with a 
velocity 
$v<w$ gain energy from the 
binary which then hardens. Likewise, stars with $v>w$ are likely to lose energy 
to the binary which then becomes wider. This is because for $v<w$, the velocity 
in the binary orbit 
is higher than the approach velocity of the incoming star and the star is more 
easily captured 
onto a bound orbit. On a bound orbit the star will eventually pass close to 
either 
one or both of the SMBHs and is then ejected from the vicinity of the SMBH 
binary. 
Faster stars with $v>w$ are less likely to become bound to the binary 
\citep{Valtonen2006}. Note that $w$ is obviously time-dependent: if the SMBH 
binary hardens, $w$ increases. Thus, there are more particles with $v<w$ to 
efficiently harden the binary. However, this effect is somewhat balanced by the 
fact that the region of strong three-body interactions ($r<6\times a$) becomes 
smaller as the SMBH binary hardens.

The velocity distributions $f(v)$ and $f(v/w)$ of the stellar particles 
crossing a shell at a distance of $30$ pc from the SMBH binary during 
$200$ Myr $<t<400$ Myr are shown in Fig. \ref{fig: velocities}.
The bulge in the H5 sample is situated in a massive 
DM halo resulting in higher velocities of the incoming stellar particles than 
in 
the bulge-only samples. Consequently, the sample H5 has the highest fraction of 
particles with velocities above the watershed velocity $w$. This is why the 
hardening 
rates of the SMBH binaries in the H5 sample are smaller than in samples B1 and 
B5, even though there is more stellar mass in sample H5 interacting with the 
binary, as can be seen in Fig. \ref{fig: rperi}.

The velocity distributions 
$f(v)$ for the bulge-only samples B1 and B5 differ at low velocities: the 
low-velocity 
tail for B1 extends to lower velocities than for sample B5. The 
reason for the differences of the hardening rates between samples B1 and B5 
becomes 
more clear when studying the incoming stellar velocity distribution $f(v/w)$ 
with respect to the time-dependent watershed velocity $w$. There is a larger 
fraction of stellar particles with velocities $v<w$ in sample B1 resulting 
in a higher hardening rate than in sample B5. In addition, there are more 
particles with small pericenter distances as seen in Fig. \ref{fig: rperi}. We 
note that the the distribution $f(v/w)$ acts as an exaggerated version of 
$f(v)$: as there are more low-velocity stellar particles in B1, the SMBH binary 
hardens faster. Thus, the watershed velocity $w$ increases, and more stellar 
particles now have $v<w$ even if $f(v)$ remains constant.

To sum up the results: we studied the distributions of pericenter 
distances and the approach velocities of the stellar particles interacting with 
the SMBH binary in order to explain why the hardening rates decrease when 
adding a massive DM halo or when increasing the stellar mass resolution. We 
find 
that sample B1 has 
both the smallest stellar approach velocities with respect to the SMBH binary 
watershed velocity $w$ and more particles at small pericenter distances from 
the 
binary. 
In the sample B5 there are slightly less particles with small pericenter 
distances and, more importantly, less particles with low velocities. The 
halo-including sample H5 has far fewer stellar particles at velocities $v<w$, 
which explains the lowest hardening rate, even though more stellar mass 
interacts with the SMBH binary for this simulation sample.
We find that this picture is consistent with the study of  
\cite{gualandris2016}: mergers without a DM halo result in a merger remnant in 
which the SMBH binary evolution depends on the particle number used when 
$N<10^7$. However, we note that while \cite{khan2011} and \cite{Preto2011} 
reported resolution-independent SMBH binary hardening in mergers without a DM 
halo, we find resolution-independent hardening rates only in the merger 
simulations, which use multi-component ICs including a DM 
halo. However, some of the differences might be attributed to the fact 
that the parameters of the progenitor galaxies and the merger orbits are not 
exactly 
the same in the different studies. 
Probing the full galaxy and merger orbit parameter space would require a far 
larger simulation sample than the 57 merger runs used in this work and is 
beyond 
the scope of the 
present study.

The hardening rates are also large enough to drive the binaries into the 
gravitational wave (GW) dominated regime ($a \sim 0.1$ pc for a circular SMBH 
binary) less than $\sim 1$ Gyr after the formation of the binary. As the binary 
eccentricities are high, $0.8 < e < 0.95$ in our highest-resolution simulations 
with a DM halo, 
the expected 
SMBH merger timescale for these simulations is less than a gigayear. We 
conclude 
that the 
final-parsec problem does not exist in our simulation sample of mergers of 
massive elliptical galaxies. In addition, the incorporation of the massive dark 
matter 
halo in the multi-component initial conditions reduces the 
resolution-dependence 
of the SMBH binary 
hardening rates and in practice makes the hardening rate resolution-independent 
within the range of particle numbers used in this study.

\subsection{SMBH merging timescales}\label{pn-section}

We run five additional simulations at the resolution of sample H5 
presented in Fig. \ref{fig: 
khan_hardening_H}, but this time including also Post-Newtonian corrections
in order to obtain the actual merging timescales of the SMBH binaries. This 
sample is labelled H5 PN. Only the SMBH-SMBH terms in the PN equations of 
motion are included since the star-SMBH 
terms would be unphysically large as the stellar particle masses are $m_\star 
\gg M_\odot$ in our simulations. We include Post-Newtonian
terms up to order PN2.5, the highest of which is the radiation-reaction term. 
We chose not to include the higher-order terms as there is still some ambiguity 
in the literature over the derivation of these terms and as 
the contribution of the next radiation-reaction term at PN3.5 is proportional 
to $\propto c^{-7}$, its effect would be anyhow negligible (see e.g. \citealt{Will2006} for further details).

\begin{figure*}
\includegraphics[width=\textwidth]{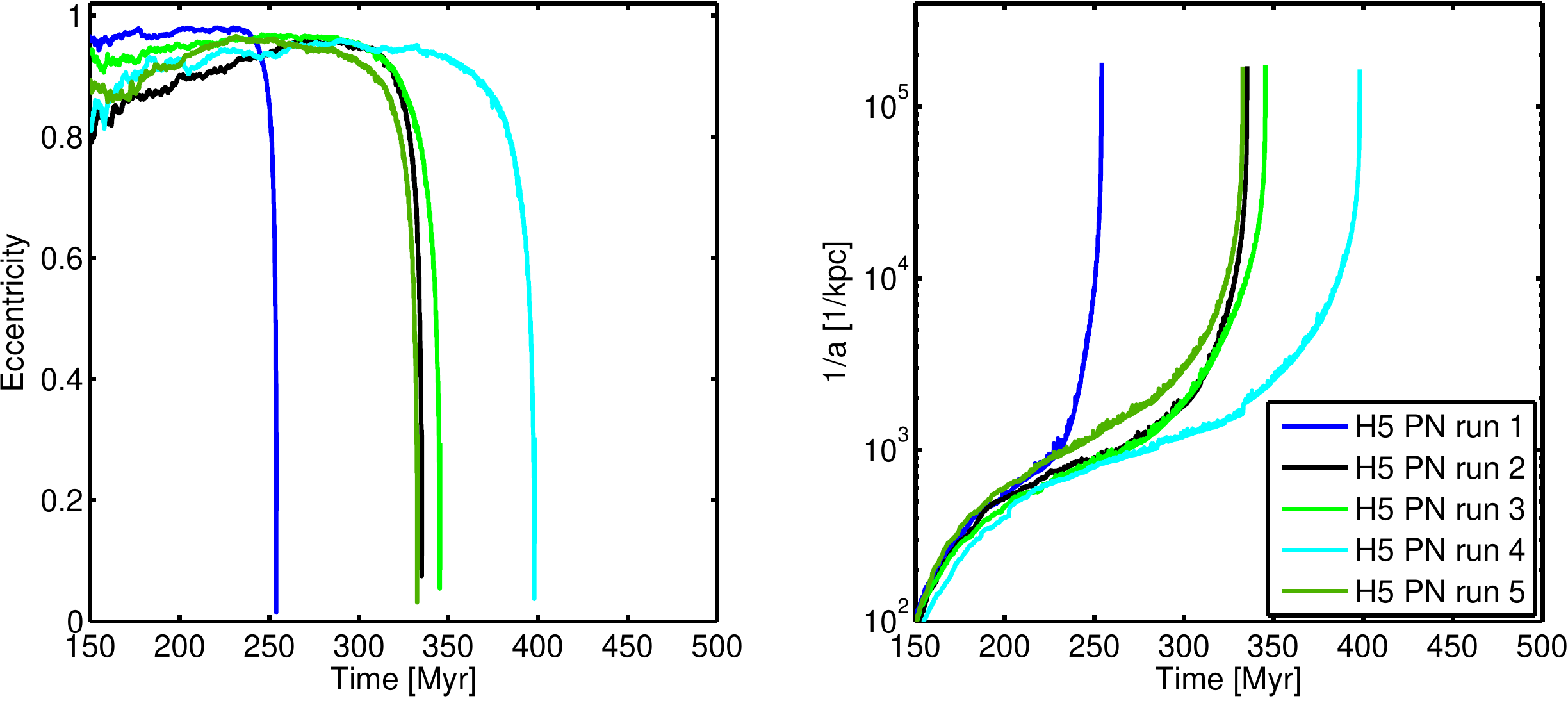}
\caption{The Post-Newtonian binary evolution in five runs from the 
simulation sample H5 PN. As the SMBH binary eccentricities are high, the SMBH mergers 
occur rapidly, $120$ Myr - $264$ Myr after the formation of the binaries. It is 
important to note that a scatter of $\Delta e \sim 0.15$ in binary eccentricity 
results in a large difference $\Delta t \sim 144 \ \rm Myr$ in the coalescence times. 
The large scatter in the SMBH coalescence timescale is due to the very steep dependence of the 
GW emission on the binary eccentricity.}
\label{fig: pn}
\end{figure*}

We run the five simulations until the coalescence of the binaries. 
The orbital evolution of the binaries is presented in Fig. \ref{fig: pn}. The 
binary evolution is qualitatively similar compared to the purely Newtonian case 
when $a > 0.75$ pc - $1$ pc: the binary hardens at a constant rate and becomes 
slowly more eccentric due to interactions with the surrounding stars. After 
this point, the GW emission determines the orbital evolution of the SMBH binary up 
to the final coalescence.

The radiative losses of binary orbital energy and angular momentum are both 
extremely sensitive to the eccentricity of the binary. During a single orbital 
period, the average PN2.5 order loss effects are, according to 
\citet{peters1963},
\begin{align}\label{eq: peters2}
-\langle \dot{E}\rangle = \frac{32 G^4 M_{\bullet,1}^2 M_{\bullet,2}^2 
(M_{\bullet,1}+M_{\bullet,2}) }{5 c^5 a^5} \frac{1+\frac{73}{24} e^2 + 
\frac{37}{96} 
e^4}{(1-e^2)^{7/2}}, \\
     -\langle \dot{L}\rangle = \frac{32 G^{7/2} M_{\bullet,1}^2 M_{\bullet,2}^2 
\sqrt{M_{\bullet,1}+M_{\bullet,2}} }{5 c^5 a^{7/2}} \frac{1+\frac{7}{8} 
e^2}{(1-e^2)^2},
\end{align}
which both diverge as $e \rightarrow 1$. For example, a binary with $e=0.95$ 
radiates $\sim 127$ times more GW energy during a single orbital period than a 
binary with $e=0.8$ assuming equal semi-major axes. For this reason, the 
initially more eccentric binaries
coalesce on much shorter timescales. At $t=150$ Myr, or $16$ Myr after 
the 
binary formation, the eccentricities of the binaries are $0.80 \lesssim e 
\lesssim 0.96$. The peak eccentricities reach $e>0.95$ for all the five 
binaries 
before the gravitational wave driven circularization phase 
begins. During the final tens of millions of years before 
the coalescence, the binary circularizes and hardens very rapidly. Both the 
orbital decay and circularization of the binary are consistent with Peters' 
analytic formulas \citep{peters1963}. The SMBH coalescences occur at 
$120$ Myr, $199$ Myr, $201$ Myr, $211$ Myr and $264$ Mr after the binary 
formation. The difference in the final BH coalescence time can vary up to 
$\sim 50$\% between the mergers, although the simulations 
were
set up with identical initial conditions, with the only difference being in the 
initial random seed.

The final SMBH separation before the coalescence is $\sim 1000$ AU in all three 
simulations. This is a significant improvement over the merging criterion 
typically used in softened \gadget{}-based codes in which the typical SMBH 
merger separation is 
of the order of $10$-$100$ pc at the time of the merger (e.g. 
\citealt{springel2005,Johansson2009a,Salcido2016}).
Obtaining the distribution of the SMBH merger timescales 
for a representative simulation sample would require substantial further work 
and is beyond the scope of this study.
However, we emphasize that the only difference in the five initial conditions 
of the presented Post-Newtonian simulations is the random number seed used in 
our multi-component initial conditions generator. The realized eccentricity 
scatter of $\Delta e \sim 0.15$ resulted in a difference of $\Delta t \sim 144$ Myr in 
the SMBH merger timescale.

Based on our three-body experiments we expect that 
the eccentricity scatter will decrease with an increasing stellar mass 
resolution. Thus, future work aiming at obtaining accurate SMBH merger 
timescales should focus on simulating the galaxy merger and the merger remnant 
in the most accurate manner and with the highest resolution possible, as the 
binary eccentricity evolution depends sensitively on the stellar velocity field 
around the binary. This ambitious goal requires going to particle numbers significantly beyond the 
value of $N_\star > 3.6 \times 10^6$ used in the present study.\\

As the hardening rate of an equal-mass SMBH binary in relatively unsensitive to
the eccentricity \citep{mikkola1992, quinlan1996}, we can estimate the maximum 
SMBH coalescence timescale for our galaxy models and SMBH masses used without 
running 
additional simulations. The maximum coalescence timescale is obtained in the 
situation in which the SMBH binary eccentricity remains zero during the entire 
slingshot-hardening phase and the GW inspiral. The merging timescale can be 
estimated as follows. Assuming that the hardening rate $s = d(1/a)/dt$ from the 
three-body interactions is constant and $e=0$, the time $t_\mathrm{f}$ from 
the initial semi-major axis $a_0$ to the final value $a_\mathrm{f}$ is given by
\begin{equation}  
t_{\mathrm{f}} = \int_{a_\mathrm{f}}^{a_0} \frac{da}{s a^2 + 
\mathrm{C}_{\mathrm{PM}} a^{-3} },
\end{equation} 
which is most conveniently solved numerically. Here 
$\mathrm{C}_{\mathrm{PM}}$ is the absolute value of the constant appearing in 
the Peters' formula (Eq. \ref{eq: peters}). Setting $a_\mathrm{f} = 1000$ AU 
mimicking the SMBH coalescence distance in \ketju{} and $a_0 = 0.01$ kpc, we 
get 
the maximum possible SMBH merging timescales of $\sim 2$ Gyr for the B5 sample  
and $\sim 3.1$ Gyr for the H5 sample using the hardening rates obtained in \S 
\ref{binaryevolution}. As mentioned in \S \ref{galaxy_models}, the DM content of 
galaxy models used in this study brackets the realistic 
population of elliptical galaxies so the maximum possible SMBH merger timescale 
is 2-3 Gyr for the used SMBH mass and the initial stellar distribution. The 
maximum merging timescales are well below the Hubble time, but are long enough 
that subsequent galaxy mergers may have time to bring in additional SMBHs near 
the 
original binary. This usually results in the rapid merging of two of the SMBHs 
while the third one is dynamically ejected from the center of the galaxy. 
Finally, we note that the 
maximum SMBH merging timescales obtained here are very likely to be dependent on 
the central stellar 
density profiles of the progenitor galaxies and may thus not be generalized in a 
straightforward manner.\\
\section{Discussion}
\label{discussion}

The new code developed in this study allows us to simultaneously model 
accurately both the dynamics close to SMBHs and
the global galactic dynamics of their host galaxies. The primary motivation for 
developing this new code was two-fold. Firstly, we want to push the numerical 
resolution 
of the galactic scale simulations ultimately to unprecedented particle numbers 
$(N\gtrsim 
10^{7}-10^{8})$ and secondly, 
the new code allows for galactic-scale dynamical studies that include also a 
gaseous component and their associated feedback processes. Both of these goals 
are difficult to 
reach with traditional $N$-body codes even allowing for GPU acceleration, as 
for 
these 
codes the particle numbers are typically limited to $N\sim 10^{6}$ and the 
inclusion of 
the gaseous component is in general not straightforward to implement (see e.g. 
\citealt{Nitadori2012,wang2014,Wang2016}).

As in the hierarchical model galaxies grow through mergers, situations with 
multiple SMBHs in the merger remnant
will be frequently encountered. Recently an increasing number of systems with 
multiple SMBHs have been detected
(e.g. \citealt{Valtonen2008, Boroson2009, Deane2014}). However, the absolute 
number of such systems still remains 
relatively low, which may indicate that the dynamical friction driven decay of 
the 
SMBH orbits and their eventual mergers 
are relatively rapid processes. 

The key prediction to come out of an accurate dynamical simulation is 
the actual merging timescale of the SMBHs. Often the orbital decay of SMBHs has 
been studied using either global isolated merger
simulations or full cosmological simulations (e.g. 
\citealt{springel2005,Sijacki2015}), which both have 
typically employed softened gravity, such as 
is used in standard \gadget{}. We showed in Fig. \ref{fig: friction} 
the well-known result that softened calculations underestimate the 
dynamical friction effect of stars with impact parameters smaller than the 
gravitational softening length, which 
depending on the situation can result in dynamical friction timescales that are 
off by a factor of a few. 
This problem is largely circumvented in direct $N$-body simulations with no or 
very low gravitational softening, 
albeit at the cost of the simulation typically starting from rather idealized 
initial conditions in which the SMBHs are already
relatively close, thus often missing the preceding dynamical friction phase 
altogether (e.g. \citealt{berczik2006,khan2011}). The 
 \ketju \ code is an optimal tool in circumventing both of these problems and 
we have demonstrated in this article that the
code can be used to study the orbital decay of SMBHs in a realistic galactic 
setting. 

The dynamical friction phase is followed by first the phase in which the binary 
hardens through three-body interactions
with the stellar background leading ultimately to the gravitational wave driven 
phase, which is responsible 
for the final merging of the SMBHs. Typically the softening length is set to 
$\sim 1-20$ pc in isolated merger 
simulations (e.g. \citealt{Hopkins2013,Choi2014}) and to $\sim 100-500$ pc 
(e.g. \citealt{Feldmann2016,Choi2016}) in high-resolution cosmological 
simulations, both of which provide insufficient
resolution for studying the binary hardening at $1-10 \ \rm pc$ spatial scales 
and the eventual GW 
driven coalescense at centiparsec scales. The typical solution for this 
predicament is the adoption of subresolution
models for the unresolved dynamics of the system, with the most 
famous prescription being the original
recipe used in \citet{springel2005}, which postulated instantaneous merging of 
nearby SMBHs, with the resulting merger
taking place typically at $\sim 10-100$ pc distances depending somewhat on the 
exact numerical resolution.   

As these subresolution descriptions lead often to very rapid SMBH merging (e.g. 
\citealt{Sijacki2015}) the 
burning question now remains whether this is realistic. The simulations 
presented in this study show that the 
final GW phase of the SMBH merger sensitively 
depends on the eccentricity of the SMBH binary, with eccentricity 
differences ($e\sim 0.80-0.96$) 
resulting in coalescense time differences of up to $\Delta t\sim 144 \ \rm Myr$ 
(see Fig. \ref{fig: pn}). This is readily expected as the energy emitted in 
gravitational waves very strongly 
depends on the eccentricity of the SMBH binary (see Eq. \eqref{eq: peters2}). 
The 
eccentricities being of paramount 
importance it is critical that the global evolution of the galaxy merger 
leading 
up to the formation of the SMBH
binary is modeled accurately, which the \ketju{} code is able to do as 
demonstrated in \S \ref{results}. 

The SMBH binaries formed during galaxy mergers are also expected to contribute 
a 
cosmological background 
of nanohertz-frequency gravitational waves. However, 
contrary to this expectation, pulsar timing array measurements during the last 
few years have to date 
yielded no sign of the expected gravitational wave background 
\citep{Shannon2015, Arzoumanian2016}. The measurements have placed an upper 
limit of $\sim 1.0 \times 10^{-15}$ for 
the characteristic strain amplitude of 
the gravitational wave
background, measured at the frequency of $f = 1 \ \rm yr^{-1}$. Currently these 
upper limits are 
inconsistent on the $\sim 1-2$ sigma level with recent models of gravitational 
wave background formation from SMBH binaries in 
galaxy mergers (e.g. \citealt{Sesana2013,McWilliams2014,Kulier2015}), 
indicating that current models of SMBH dynamics might overestimate the SMBH 
merger rates, or that the SMBH masses themselves may be systematically 
overestimated (e.g. \citealt{Rasskazov2016, Sesana2017}). 
Thus it is important to provide more accurate theoretical predictions of the 
SMBH merger timescales that might
help in reconciling the derived upper limits of the pulsar timing array 
measurements with theory and ultimately pave
the way for more accurate predictions required by future space-borne GW 
detection experiments (e.g. \citealt{Amaro-Seoane2012}).

We have here demonstrated the functionality of \ketju{} in collisionless 
simulations that include SMBHs and both stellar 
and dark matter particles. However the real future strength of the code lies in 
realistic galactic-scale simulations that
also include a gaseous component. In the \ketju{} code the \archain{} algorithm
solves the stellar dynamics close to the SMBH, whereas all of the astrophysical 
processes, such as hydrodynamics, 
cooling, star formation and stellar feedback is performed in the \gadget{} code 
using already existing routines 
\citep{Hu2014,Hu2016}. Unlike the collisionless stellar 
component collisional, gas can cool 
and collapse towards the center of the galaxy, thus shaping the central 
gravitational potential with potentially interesting consequences for the
final-parsec problem (e.g. \citealt{Lodato2009}). However, special care will be 
required for the treatment of feedback from the SMBHs, as we can now 
dynamically 
probe 
centiparsec scales for which the standard approach of taking the Bondi-Hoyle 
accretion rate is destined to fail (e.g. \citealt{Curtis2015}).

In addition, simulations that accurately obtain the coalescence timescale of 
binary SMBHs in gas-rich galaxy 
mergers are also essential for providing accurate estimates for the expected 
recoil velocities of merged SMBHs and the
likelihood that the recoiling SMBHs escape the deepening central gravitational 
potential (e.g. \citealt{Blecha2011,Blecha2016}). 
The recoil probability will be particularly relevant for large-scale 
cosmological simulations such as Illustris and EAGLE 
\citep{Vogelsberger2014,Schaye2015} that typically assume rapid coalescense 
below the 
spatial resolution limit and which also often fix the SMBHs to the center of 
the 
potential using an artificial 
repositioning scheme \citep{Johansson2009a}. The more accurately predicted 
coalescense times of SMBHs in cosmological simulations  
will also directly impact the estimate of the strength of the gravitational 
wave background. Finally, the accurate dynamical modeling
of the SMBHs could shed light on the possible offsets between the central SMBH 
and the hosting dark matter halo, which
will be important for experiments that attempt a direct detection of the 
possible central dark matter annihilation signal (e.g. 
\citealt{Batcheldor2010,Lacroix2014}). SMBH binaries can also change the 
stellar dynamics of the nuclear regions of the host galaxy, which can be now 
probed with integral field unit (IFU) observations \citep{thomas2016}.

\section{Conclusions}
\label{conclusions}

In this article we present and test the performance of \ketju{}, a new 
regularized 
tree code based on 
algorithmic chain regularization (\archain{} integrator) implemented into 
\gadget{} \citep{springel2005b}. The key feature of the \ketju{} code is the 
inclusion of regularized regions around every SMBH, which
allows for the accurate simultaneous modeling of the global dynamics on 
galactic scales together with
the sub-parsec-scale SMBH evolution. The code also includes Post-Newtonian 
corrections up to 
order PN3.5, which also includes the PN2.5 term responsible for gravitational 
wave emission. This allows us 
in principle to follow the evolution of the SMBH binary down to spatial scales 
of $\sim 10$ Schwarzschild radii, 
which is an improvement of several orders of magnitude over the merger 
separations traditionally employed 
in softened \gadget-based codes ($\sim 10-100 \ \rm pc$, e.g. 
\citealt{springel2005,Johansson2009a,Salcido2016}).
The more accurate treatment of all three phases (dynamical friction, three-body 
interactions and GW emission) 
relevant for SMBH merging in one global simulation provides improved 
constraints 
on the SMBH coalescense timescales 
in galaxy mergers with potentially important implications for future 
gravitational wave experiments, such as LISA (e.g. \citealt{Amaro-Seoane2007}).

The  \ketju{} code was calibrated against NBODY7 
\citep{aarseth2012} by running test simulations. In the first simulation 
we followed the inspiral of a single SMBH in a Hernquist sphere, whereas in the 
second simulation we studied the hardening of a SMBH binary system
inside another Hernquist sphere. In both tests we were able to reproduce the 
results of NBODY7 at high accuracy thus validating the performance
of the \archain{} integrator and the \ketju{} code. In addition the 
energy conservation was demonstrated to be on a good level both for the 
stand-alone
\archain{} integrator and the \ketju{} code, which performed slightly better 
than the standard \gadget{} code. The scaling of the \ketju{} code was 
adequate 
up 
to $\sim 75$ processors. We conclude that this is an area which will 
require 
substantial future improvement, if the ultimate goal of simulating up to 
$N\sim 10^{7}-10^{8}$ particles is to be reached. 

We used the \ketju{} code to study the resolution dependence of the SMBH binary 
hardening rates and the SMBH coalescense timescales in a set of 
galaxy mergers. We extend the initial conditions of the progenitor galaxies to 
include in addition to the stellar bulge and SMBH, also a dark matter
halo component motivated by the studies of \citet{hilz2012}. We run a total of 
57 merger simulations with particle numbers reaching up to 
$2\times10^6$ DM particles and $2\times10^{6.25} \approx 3.6\times10^6$ stellar 
particles for the highest resolution simulations. Half of the simulations 
included only a bulge component in addition to the SMBH, whereas the other half 
also included a dark matter component. 
We found a mild dependence of the SMBH binary hardening rate on the 
mass 
resolution of 
the simulation in the simulations without a DM halo. The time evolution of the 
inverse semi-major axis of the binary orbit was proportional to particle number 
$N$ as $d(1/a)/dt \propto N^{-0.18 \pm 0.04}$ for the mergers without 
a DM halo. In the halo-including sample we found $d(1/a)/dt \propto N^{-0.03 
\pm 0.06}$, which is consistent with no resolution-dependence. We thus 
see a weaker dependence on numerical resolution for the simulation sample 
that includes a massive DM halo. We did not encounter the final-parsec problem in any of our merger simulations. 
This result is attributed to the non-spherical structure of the merger 
remnants. The non-spherical shape of the merger remnant torques the stellar orbits to fill 
the SMBH binary loss cone faster than the two-body 
relaxation, which is artificially boosted by the low mass resolution (e.g. 
\citealt{berczik2006}). We find that this result is in good agreement with 
the previous work of \citet{gualandris2016}. 

The shapes of the stellar components of the merger remnants are very 
similar in all simulations including the DM component. We attribute this to the 
fact that the massive halo dominates the galactic potential and thus relaxation 
effects in the stellar component are less important. This results in similar 
distributions of pericenter distances and approach velocities for the stars 
interacting with the SMBH binary, and consequently to nearly similar SMBH 
binary hardening rates at all resolutions. In the bulge-only mergers the differences 
in the merger remnant shapes are larger (see Fig. \ref{fig: khan_scaling}) as the 
relaxation effects are more important in the absence of the massive DM halo. In 
low-resolution bulge-only simulations, there are more stellar particles with 
low velocities and small pericenter distances compared to the high-resolution 
bulge-only 
simulations. This explains the resolution-dependence of hardening rates for the 
bulge-only simulation sample. In addition, the SMBH binary hardening rates are 
higher in 
bulge-only simulations compared to the hardening rates of simulations that 
include a massive DM halo, due to the higher stellar velocities in these 
galaxies.
The eccentricities of the SMBH binaries formed in the merger remnants are in 
general very high ($e \sim 0.9$), which is a direct consequence of the 
initial
orbits of the progenitor galaxies. However, we see a scatter in the 
eccentricity of $\Delta e \sim 0.15$ for simulation setups that only differ in 
the random 
seed used to generate the multi-component initial conditions. The scatter in 
eccentricity produced essentially by stochastic encounters between the SMBH
and stellar particles \citep{mikkola1992} can result in considerable 
differences in the SMBH coalescence timescales of the order of $\sim 140 \ \rm 
Myr$ due to the steep scaling of gravitational wave emission with the binary 
eccentricity (see Eq. \eqref{eq: peters2}).

Development of hybrid regularized tree codes that allow the simultaneous 
modeling of the global dynamics on galactic scales together with
the sub-parsec-scale SMBH evolution will hopefully help to bridge the gap 
between the galaxy formation and $N$-body communities. We find the \ketju{} code
is an important step in this direction, but we end by stressing that the 
particle and the stellar mass resolutions significantly beyond the ones 
employed in this study are still required for obtaining truly highly accurate 
predictions for the timescales of SMBH mergers.

\small
\begin{acknowledgements}
The authors thank the anonymous referee for constructive comments which helped 
to improve the manuscript. A. R., P. H. J., P. P. and T. N. wish to thank Dr. 
Simon Karl and Dr. Sverre 
Aarseth for assistance related to regularized chain calculations and N-body 
simulations. The numerical simulations were performed on facilities hosted by 
the 
CSC -IT Center for Science in Espoo, Finland. A. R., P. H. J., P. P. and N. L. 
acknowledge support from the MPA Garching Visitor Programme. A. R. is funded by 
the 
doctoral programme of Particle Physics and Universe Sciences at the University 
of Helsinki. P. H. J., 
P. P., N. L. T. S. acknowledge the support of the Academy of Finland grant 
1274931. N. L. is 
supported by the Jenny and Antti Wihuri Foundation. T. N. acknowledges support 
from the 
DFG Cluster of Excellence 'Origin and Structure of the Universe'.

\end{acknowledgements}

\normalsize
\newpage

\bibliography{references}

\newpage

\appendix

\section{Appendix A: Algorithmic regularization}\label{sc:algoreg}
We start with the usual Newtonian N-body Hamiltonian
\begin{equation}
    H = \sum_i \frac{1}{2}m_i\norm{\vect{v}_i}^2 - \sum_{i}\sum_{j>i}\frac{Gm_i
    m_j}{\norm{\vect{r}_{ij}}} = T - U,
\end{equation}
where $m_i$ are the masses, $\vect{v}_i$ velocities and $\vect{r}_{ij} =
\vect{r}_j-\vect{r}_i$ relative separations of the N bodies, $T$ is the
total kinetic energy and $U$ is the force function, or 
equivalently the negative of the Newtonian potential energy. 
The binding energy of the system is $B=-H = U - T$.
This Hamiltonian generates the usual N-body equations of motion
\begin{align}
    \dot{\vect{r}}_i &= \vect{v}_i\\
    \dot{\vect{v}}_i &= \sum_{j\neq i}
    \frac{Gm_j\vect{r}_{ij}}{\norm{\vect{r}_{ij}}^3}.
\end{align}
This system is transformed by introducing a new fictitious time, or independent
variable, $s$ with the definitions
\begin{equation}\label{eq:ttl}
    \ud s = \left[\alpha(T + B) + \beta\omega +\gamma\right]\ud t
    = \left[\alpha U + \beta\Omega +\gamma\right]\ud t,
\end{equation}
where $\alpha$, $\beta$ and $\gamma$ are real constants,
$\Omega(\{\vect{r}_{i}\})$ is in principle an arbitrary real-valued function of 
the
coordinates, and $\omega$ is a new velocity-like variable, for which we
define
\begin{equation}
    \dot{\omega} = \sum_i \boldsymbol{\nabla}_{\vect{r}_i}\Omega\cdot\vect{v}_i
\end{equation}
and $\omega(0) = \Omega(\{\vect{r}_{i}(0)\})$.
Following \citet{mikkola2006,mikkola2008} we set
\begin{equation}
    \Omega = \sum_{i<j} \frac{\Omega_{ij}}{\norm{\vect{r}_{ij}}},
\end{equation}
where $\Omega_{ij}=\tilde{m}^2$ if $m_i m_j < \epsilon_\Omega
\tilde{m}^2$ and $\Omega_{ij}=0$ otherwise. 
Here
\begin{equation}
    \tilde{m}^2 = \frac{2}{N(N-1)}\sum_{i<j} m_i m_j,
\end{equation}
is the mean product of particle masses, with $N$ equal to the number of
particles. Following \citet{mikkola2008}, we set the parameter
$\epsilon_\Omega=10^{-3}$. This choice of $\Omega$ is made to guarantee that
low-mass particles have a non-negligible effect on the regularization,
even though they make only a negligible contribution to the value of $U$.

With these definitions, the two definitions in Eq. \eqref{eq:ttl} are
equivalent on the exact integral curve, since $T+B=U$ and
$\Omega(\{\vect{r}_i(t)\}) = \omega(t)$.
With prime signifying a derivative with respect to $s$, we get
the following equations of motion
for the coordinates,
\begin{align}
    t' &= \left[\alpha(T + B) + \beta\omega +\gamma\right]^{-1} \\
    \vect{r}'_i &= t'\vect{v}_i
\end{align}
and for the velocities
\begin{align}
    t' &= \left(\alpha U + \beta\Omega +\gamma\right)^{-1} \\
    \omega' &= t' \sum_i\boldsymbol{\nabla}_{\vect{r}_i}\Omega\cdot\vect{v}_i \\
    \vect{v}'_i &= t' \sum_{j\neq i}
\frac{Gm_j\vect{r}_{ij}}{\norm{\vect{r}_{ij}}^3}.
\end{align}
In the presence of perturbing accelerations $\vect{f}_i$, the
binding energy of the system will not be constant, but instead
\begin{equation}
    \dot{B} = -\dot{T} + \dot{U} =-\sum_i m_i \vect{v}_i\cdot\vect{f}_i,
\end{equation}
since the Keplerian accelerations cancel.
With this addition the time transformed velocity equations read
\begin{align}
    t' &= \left(\alpha U + \beta\Omega +\gamma\right)^{-1} \\
    B' &= -t' \sum_i m_i \vect{v}_i\cdot\vect{f}_i \\
    \omega' &= t' \sum_i\boldsymbol{\nabla}_{\vect{r}_i}\Omega\cdot\vect{v}_i \\
    \vect{v}'_i &= t'\left(
        \sum_{j\neq i} \frac{Gm_j\vect{r}_{ij}}{\norm{\vect{r}_{ij}}^3}
        + \vect{f}_i
    \right).
\end{align}
Different choices of the
parameter triplet $(\alpha,\beta,\gamma)$ correspond to different
algorithmic regularization schemes, with $(1,0,0)$ yielding the
logarithmic Hamiltonian method \citep{mikkola1999,preto1999} and
$(0,1,0)$ being equivalent to the Time-Transformed Leapfrog scheme
\citep{mikkola2002}. Both give exact orbits for two-body orbits
including collisions. We choose
$(\alpha,\beta,\gamma)=(1,0,0)$, based on the suggestion in
\citet{mikkola2008}.
For a more extended discussion of possible parameter choices,
see \citet{mikkola2006} and \citet{mikkola2008}. 

\section{Appendix B: Chained leapfrog}\label{sc:chain_leapfrog}
\label{appendix_B}

When the Eqs. \eqref{eq:chaineqs} are time
transformed as described in Appendix~\ref{sc:algoreg}, 
we end up with the following equations of motion
for the chain coordinates
\begin{align}
    t' &= \left[\alpha(T + B) + \beta\omega +\gamma\right]^{-1} \\
    \vect{X}'_i &= t' \vect{V}_i,
\end{align}
and the velocities
\begin{align}
    t' &= \left(\alpha U + \beta\Omega +\gamma\right)^{-1} 
\label{eq:chainveleqsa} \\
    B' &= -t' \sum_i m_i\vect{v}_i\cdot\vect{f}_i \\
    \omega' &= t' \sum_i\boldsymbol{\nabla}_{\vect{X}_i}\Omega\cdot\vect{V}_i \\
    \vect{V}'_i &= t' \left(\vect{A}_i(\{\vect{X}_i\}) + \vect{f}_i + 
\vect{g}_i(\vect{v}_i)\right)\\
    \vect{S}'_i &= t' \vect{S}_{\text{PN},i} \times \vect{S}_i, 
\label{eq:chainveleqsb}\\
\end{align}
where $\vect{g}_i(\vect{v}_i)$ are the perturbing accelerations which
depend on the particle velocities.
If the perturbing accelerations do not depend on the particle
velocities, that is $\vect{g}_i=0$,
the Eqs. \eqref{eq:chainveleqsa}-\eqref{eq:chainveleqsb}
can immediately be integrated with a standard leapfrog. 
In this case, while the equations of motion for $B$ and $\omega$ depend
on velocities, the dependence is linear, and as such can be analytically
integrated over one timestep. It should be noted that in practice it is
easier to evaluate the derivatives of $B$ and $\omega$ using the
non-chained velocities $\vect{v}_i$ and coordinates $\vect{r}_i$, but using 
Eq. \eqref{eq:chaindist} for the relative distances $\vect{r}_{ij}$.

However, it should be noted that the Post--Newtonian corrections
(\S~\ref{sc:PN}) do depend on
the particle velocities, and possibly spins, in addition to the particle
coordinates. In this case, the equations of motion
\eqref{eq:chainveleqsa}-\eqref{eq:chainveleqsb} are
no longer integrable, and normally an implicit method would have to be
used to compute the solution over one timestep.
This is undesirable, since an implicit method generally requires
iterating to convergence, which would force re-evaluating the
computationally heavy PN
corrections several times. This can be avoided by extending the phase
space, which allows an explicit leapfrog to be constructed, as in
\citet{hellstrom2010} or using the generalization derived in 
\citet{pihajoki2015}. 
Combined with the extrapolation method, this allows the PN corrections
to be implemented with a much smaller computational overhead.

The phase space extension is done by introducing auxiliary velocities
$\vect{w}_i$ and the corresponding auxiliary chained velocities $\vect{W}_i$ as 
well
as auxiliary spins $\vect{Z}_i$ for each particle. An auxiliary 
time transformation variable $\sigma$ is required as well.
Before the leapfrog
step, these are set to the values of the original variables. Then,
when the velocity kicks are calculated, the value of $\vect{w}_i$ is 
used to compute $\dot{\vect{v}}_i$, and similarly the value of
$\vect{v}_i$ is used to compute $\dot{\vect{w}}_i$. Evolution of the spin
and the time transformation variable is done in a similar alternating way.

The end result is a leapfrog, which can be written with mappings
$\chdrift(\Delta s)$ (drift) and $\chkick(\Delta s)$ (kick), which
propagate the system in the phase space. We list the algorithmic form of
these mappings below.
The maps can then be combined to form the two second order leapfrogs
commonly called drift-kick-drift (DKD)
leapfrog $\chdrift(h/2)\chkick(h)\chdrift(h/2)$ and the kick-drift-kick (KDK)
leapfrog $\chkick(h/2)\chdrift(h)\chkick(h/2)$, where $h$ is the timestep.
In the code, the DKD leapfrog is used.

\newpage

\begin{algorithm}[H]
\begin{algorithmic}
    \Procedure{$\chdrift$}{$\Delta s$}
    \State $T \gets \sum_{i=1}^{N} \frac{1}{2} m_i \norm{\vect{v}_i}^2$
    \State $\Delta t \gets \Delta s / \left[\alpha (T+B) + \beta\omega 
+\gamma\right]$
    \State $t \gets t + \Delta t$
    \For{$i\gets1,\ldots,N-1$}
        \State $\vect{X}_i \gets \vect{X}_i + \Delta t\, \vect{V}_i$
    \EndFor
    \EndProcedure
\end{algorithmic}
   \caption{$\chdrift(\Delta s)$}
   \label{alg:drift}
\end{algorithm}

\begin{algorithm}[H]
\begin{algorithmic}
    \Procedure{PhysicalStep}{$\Delta t$}
    \State{Evaluate $\vect{w}_i$ using $\vect{W}_i$, $\vect{g}_i$ using 
$\vect{w}_i$ 
and $\vect{S}_{\text{PN},i}$ using $\vect{Z}_i$}
    \State $\omega \gets \omega + \Delta t\,\sum_{i=1}^N 
\boldsymbol{\nabla}_{\vect{r}_i}\Omega\cdot\vect{w}_{i}$
    \State $B \gets B - \Delta t\sum_{i=1}^N m_i \vect{w}_i\cdot (\vect{f}_i + 
    \vect{g}_i)$
    \For{$i\gets1,\ldots,N-1$}
        \State $\vect{V}_i \gets \vect{V}_i + \Delta t \left(\vect{F}_{i+1} - 
\vect{F}_{i} + \vect{f}_{i+1}-\vect{f}_i + \vect{g}_{i+1}-\vect{g}_i\right)$
    \EndFor
    \For{$i\gets1,\ldots,N$}
        \State $\vect{S}_i \gets \vect{S}_i + \Delta t 
\,\vect{S}_{\text{PN},i}\times \vect{Z}_i$
    \EndFor
    \EndProcedure    
    \Statex
    \Procedure{AuxiliaryStep}{$\Delta t$}
    \State{Evaluate $\vect{v}_i$ using $\vect{V}_i$, $\vect{g}_i$ using 
$\vect{v}_i$ 
and $\vect{S}_{\text{PN},i}$ using $\vect{S}_i$}
    \State $\sigma \gets \sigma + \Delta t\sum_{i=1}^N 
\boldsymbol{\nabla}_{\vect{r}_i}\Omega\cdot\vect{v}_{i}$
    \For{$i\gets1,\ldots,N-1$}
        \State $\vect{W}_i \gets \vect{W}_i + \Delta t \left(\vect{F}_{i+1} - 
\vect{F}_{i} + \vect{f}_{i+1}-\vect{f}_i + \vect{g}_{i+1}-\vect{g}_i\right)$
    \EndFor
    \For{$i\gets1,\ldots,N$}
        \State $\vect{Z}_i \gets \vect{Z}_i + \Delta t 
\,\vect{S}_{\text{PN},i}\times \vect{S}_i$
    \EndFor
    \EndProcedure
    \Statex    
    \Procedure{$\chkick$}{$\Delta s$}
    \State $U \gets \sum_{i=1}^N\sum_{j>i}^N \frac{Gm_i m_j}{r_{ij}}$
    \State $\Omega \gets \Omega(\vect{r}_{1},\ldots,\vect{r}_N)$
    \State $\Delta t \gets \Delta s / \left(\alpha U + \beta\Omega 
+\gamma\right)$
    \For{$i\gets1,\ldots,N$}
        \State $F_i = \sum_{j\neq i} 
\frac{Gm_j\vect{r}_{ij}}{\norm{\vect{r}_{ij}}^3}$
    \EndFor
    \State{Compute velocity independent perturbations $\vect{f}_i$}
    \State \textsc{AuxiliaryStep}$(\Delta t/2)$
    \State \textsc{PhysicalStep}$(\Delta t)$
    \State \textsc{AuxiliaryStep}$(\Delta t/2)$
    \EndProcedure 
\end{algorithmic}
    \caption{$\chkick(\Delta s)$}
    \label{alg:kick}
\end{algorithm}

\end{document}